\documentclass[]{aa}  

\usepackage{graphicx}
\usepackage{txfonts}
\usepackage{inputenc}
\usepackage[]{hyperref}
\hypersetup{colorlinks=true, allcolors=blue}
\usepackage{multicol}
\usepackage{multirow}
\usepackage{siunitx}
\DeclareSIUnit{\solarmass}{\text{M}_{\odot}}
\DeclareSIUnit{\parsec}{pc}
\DeclareSIUnit{\year}{yr}
\DeclareSIUnit{\mag}{mag}
\DeclareSIUnit{\arcsec}{arcsec}
\usepackage{amsmath}
\usepackage{xcolor}
\usepackage{booktabs}
\usepackage{placeins}
\makeatletter
\renewcommand*\aa@pageof{, page \thepage{} of \pageref*{LastPage}}
\makeatother
\begin{document} 
   \title{The physical properties of Cluster Chains}

   \author{Laura Posch \inst{1},
          Jo\~ao Alves \inst{1},
          Núria Miret-Roig \inst{1},
          Sebastian Ratzenb\"ock \inst{1, 2},
          Josefa Gro\ss schedl \inst{3, 4, 1},
          Stefan Meingast \inst{1},
          Cameren Swiggum \inst{1},
          Ralf Konietzka \inst{5, 6}
          }

   \institute{
             University of Vienna, Department of Astrophysics, T\"urkenschanzstrasse 17, 1180      Vienna, Austria
        \and
             University of Vienna, Research Network Data Science at Uni Vienna, Kolingasse 14-16, 1090 Wien, Austria
        \and
             Czech Academy of Sciences, Boční II 1401, 141 31 Prague 4, Czech Republic
        \and
             Physikalisches Institut, Universität zu Köln, Zülpicher Str. 77, D-50937 Köln, Germany
        \and
             Center for Astrophysics $\mid$ Harvard \& Smithsonian, 60 Garden St., Cambridge, MA 02138, USA
        \and
             Department of Astronomy, Harvard University, 60 Garden St., Cambridge, MA 02138, USA
        }

   \date{Received June 29, 2024; accepted October 22, 2024}
 
    \abstract
    {
    We explore the kinematics and star formation history of the Scorpius Centaurus (Sco-Cen) OB association following the initial identification of sequential, linearly aligned chains of clusters. Building upon our characterization of the Corona Australis (CrA) chain, we now analyze two additional major cluster chains that exhibit similar characteristics: the Lower Centaurus Crux (LCC) and Upper Scorpius (Upper Sco) chains.
    All three cluster chains display distinct sequential patterns in 1) the 3D spatial distribution, 2) age, 3) velocity, and 4) mass. The Upper-Sco chain is the most massive and complex cluster chain, possibly consisting of two or more overlapping subchains. We discuss the possible formation of cluster chains and argue for a scenario where feedback from the most massive star formation episode 15~Myr ago initiated the formation of these spatio-temporal cluster sequences.
    Our results identify cluster chains as a distinct type of stellar structure with well-defined physical properties, formed in environments capable of sustaining stellar feedback over timescales of 5--10 Myr.
    We find that around 40\% of the stellar population in Sco-Cen formed due to triggered star formation, with 35\% forming along the three cluster chains.
    We conclude that cluster chains could be common structures in OB associations, particularly in regions that have similar natal environments as Sco-Cen.
    Beyond their significance for star formation and stellar feedback, they appear to be promising laboratories for chemical enrichment and the transport of elements from one generation to the next in the same star-forming region.
    }

   \keywords{Stars: kinematics and dynamics -- ISM: kinematics and dynamics -- Galaxy: open clusters and associations: individual: Scorpius-Centaurus}

   \titlerunning{The physical properties of cluster chains}
   \authorrunning{Posch et al.}

   \maketitle
   

\defcitealias{Ratzenboeck_23a_AA}{R23a}
\defcitealias{Ratzenboeck_23b_AA}{R23b}
\defcitealias{Posch_23_AA}{P23}


\section{Introduction}\label{sec:introduction}

    Stellar feedback plays a crucial role in the assembly of star-forming regions and the dispersal of the gas \citep[e.g.,][]{Elmegreen_Lada_77_ApJ, Elmegreen_11_EAS}. Winds and photoionization from massive stars and supernova (SN) explosions can profoundly transform the interstellar medium (ISM), compress the gas, and induce star formation. This is supported by both observations \citep[e.g.,][]{McCray_83, deGeus_92_AA, Megeath_96_AA, Megeath_97_AJ, Preibisch_99_AJ, Zavagno_06_AA, Thompson_12_MNRAS, Getman_12_MNRAS, Kendrew_12_ApJ, Malinen_14_AA, Deb_18_MNRAS, Marshall_19_MNRAS, Foley_23_ApJ} and theoretical models \citep[e.g.,][]{Klein_80_SSRv, Miao_06_MNRAS, Hosokawa_Inutsuka_06_ApJ, Dale_07_MNRAS, Bisbas_11_ApJ, Dale_12_MNRAS, Walch_13_MNRAS, Rogers_Pittard_13_MNRAS, Inutsuka_15_AA, Girichidis_16_MNRAS, Herrington_23_MNRAS}.

    The formation of linear cluster sequences as a result of triggered star formation has been a long-standing prediction \citep[e.g.,][]{Elmegreen_Lada_77_ApJ, Whitworth_94_AA, Elmegreen_11_EAS}.
    Previous studies have identified broad spatio-temporal gradients within HII regions using young stellar object (YSO) classifications \citep[e.g.,][]{Getman_07_ApJ, Getman_09_ApJ}. The unprecedented precision of the \textit{Gaia} satellite \citep{Gaia_16_AA}, together with the clustering accuracy of fine-tuned machine learning algorithms, now reveal well-defined sequences of young clusters in time and space with greater accuracy \citep{Ratzenboeck_23b_AA} (hereafter \citetalias{Ratzenboeck_23b_AA}).

    \begin{figure*}
        \centering
        \includegraphics[width=\textwidth]{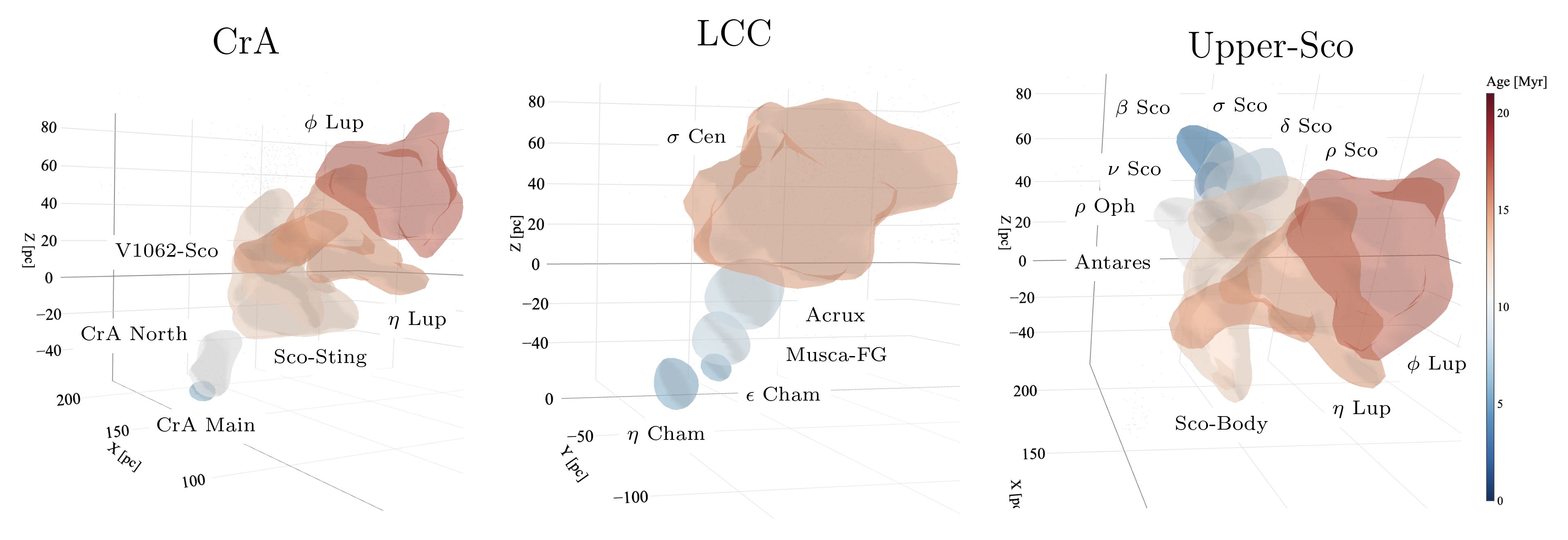}
        \caption{3D visualization of the CrA, LCC, and Upper-Sco cluster chains. The figure shows the distribution of Sco-Cen clusters colored by age (from \citetalias{Ratzenboeck_23b_AA}).
        An interactive version of the 3D positions of these cluster chains is available \protect\href{https://homepage.univie.ac.at/laura.posch/wp-content/uploads/2024/06/Plotly_ScoCen_Volumn_CHAINS.html}{here}.}
        \label{fig:sigma-chains}
    \end{figure*}
    
    The Scorpius-Centaurus OB association (Sco-Cen) \citep[e.g.,][]{Blaauw_46_PGro, Blaauw_64_ARAA, deGeus_89_AA, deZeeuw_99_AJ, Preibisch_08_book, Pecault_Mamajek_16_MNRAS, Wright_Mamajek_18_MNRAS} is the closest OB association to Earth and a well-established laboratory to study star formation processes. Historically, it was divided into three major subgroups: Upper-Centaurus Lupus (UCL), Lower Centaurus Crux (LCC), and Upper-Scorpius (Upper-Sco). Previous studies suggest that stellar feedback from UCL, initiated star formation in Upper-Sco and LCC and shaped the ISM in Lupus and Ophiuchus \citep[e.g.,][]{Preibisch_99_AJ, Gaczkowski_17_AA, Krause_18_AA}. \cite{Bouy2015-ce}, in their reanalysis of \textit{Hipparcos} data, proposed that the star-forming clouds surrounding Sco-Cen are remnants of gas left over from the formation of massive stars, later triggered into star formation by stellar feedback.
    With the advent of \textit{Gaia} \citep{Gaia_18_AA, Gaia_23_AA}, substantial spatial and kinematic substructure was found in each of the three historical groups of Sco-Cen \citep[e.g.,][]{Goldman_18_ApJ, Krause_18_AA, Damiani_19_AA, Kerr_21_ApJ}. For example, Upper-Sco was found to comprise around seven to eight clusters following a distinct age gradient, which indicates a sequential star formation history \citep[e.g.,][]{Squicciarini_21_MNRAS, MiretRoig_22_AA, Briceno_23_MNRAS}.
    
    Recently, \citetalias{Ratzenboeck_23a_AA} identified 37 clusters\footnote{We use ``cluster'' in the statistical sense of the word, meaning, an enhancement over a background \citepalias{Ratzenboeck_23a_AA}. This avoids creating a new term for the spatial/kinematical coherent stellar groups in Sco-Cen without the need to establish cluster boundness, an observationally non-trivial task. None of the \citetalias{Ratzenboeck_23a_AA} clusters, except for the few embedded ones, is expected to be gravitationally bound.} in Sco-Cen using the \texttt{SigMA} algorithm. Following this work, \cite{Ratzenboeck_23b_AA} (hereafter \citetalias{Ratzenboeck_23b_AA}) computed the ages for each of the Sco-Cen clusters, providing the largest homogeneously sampled cluster census for this star formation region. This comprehensive cluster sample allowed \citetalias{Ratzenboeck_23b_AA} and \cite{Posch_23_AA} (hereafter \citetalias{Posch_23_AA}) to uncover linear spatio-temporal sequences of young clusters. \citetalias{Posch_23_AA} linked the Corona Australis (CrA) clusters to the massive clusters at the center of Sco-Cen, forming a $\sim$\SI{100}{\parsec} long linear sequence, which they called the CrA chain of clusters. Analyzing its kinematics, they found, unexpectedly, accelerated motions along the CrA chain, with the youngest cluster, located furthest below the Galactic plane, moving the fastest away from it. \citetalias{Posch_23_AA} proposed that stellar feedback from the older massive clusters in Sco-Cen was responsible for the acceleration of the CrA clusters away from the Galactic plane. 

    In this paper, we extend the analysis to include the LCC and Upper-Sco cluster chains identified in \citetalias{Ratzenboeck_23b_AA} and investigate their physical properties. We found {further} discerning properties for the cluster chains, establishing them as a new type of stellar structure with well-defined physical properties. Given their unique configuration, cluster chains are promising laboratories for stellar feedback and chemical enrichment in OB associations. By characterizing these well-established stellar structures we aim to open a path for the identification of these structures elsewhere and enhance our understanding of star formation.
    
    The paper is structured as follows. In Sect.~\ref{sec:data}, we describe the selection process applied to the data, which was used to obtain the results outlined in Sect.~\ref{sec:results}. The results are discussed in Sect.~\ref{sec:discussion} and briefly summarized in Sect.~\ref{sec:conclusion}. Additional information on data and methods are described in the Appendices~\ref{app:prop_tables} to \ref{app:rhoOph}.


\section{Data}\label{sec:data}

    We analyzed the cluster chains using a subset (19) of the 34 clusters that \citetalias{Ratzenboeck_23b_AA} associated with Sco-Cen.
    The clusters included in this study as part of a cluster chain are shown in Fig.~\ref{fig:sigma-chains},
    which illustrates their 3D distribution and positions within each chain. The volumes are color-coded according to the cluster ages \citepalias{Ratzenboeck_23b_AA}.
    The cluster selection for each chain follows the classification scheme outlined in \citetalias{Ratzenboeck_23b_AA}, which delineates distinct regions including CrA, LCC, and Upper-Sco. For the CrA chain, we used the same clusters as \citetalias{Posch_23_AA}, where we added older clusters whose kinematic properties align with the motion of the CrA chain, extending the chain to the center of Sco-Cen. Similarly, we added $\phi$~Lupus ($\phi$~Lup) and $\eta$~Lupus ($\eta$~Lup) to the Upper-Sco chain. We excluded US-foreground from our analysis, as its age and velocity were inconsistent with the rest of the Upper-Sco chain.
    The spatial and kinematic properties for all clusters within each cluster chain are summarized in Table~\ref{tab:CrA}.
    We adopt the cluster ages from \citetalias{Ratzenboeck_23b_AA} fitted to PARSEC v1.2S \citep[hereafter Parsec;][]{Bressan_12_MNRAS, Chen_15_MNRAS, Marigo_17_ApJ} model isochrones and focus on the relative age differences between clusters.  

    \begin{table*}[]
    \centering
    \caption{Summary of the physical parameters defining the chains of clusters in Sco-Cen.}
    \label{tab:chains}
    \begin{tabular}{l|ccccccc}
    \toprule \toprule
    cluster chain & N$_{\text{stars}}$ & average acceleration       & M$_{\text{res. cloud}}$ & reference               & RV$_{\mathrm{CO}}$ & P$_{\text{res. cloud}}$     & N$_{\text{SN}}$ (2$\sigma$) \\
                  &                    & (km s$^{-1}$ Myr$^{-1}$)   & (M$_{\odot}$)           &                         & (km s$^{-1}$)      & (M$_{\odot}$ km s$^{-1}$)   &                             \\
    \midrule
    CrA           & 579  (3491)        & 0.6 $\pm$ 0.1              & 9000                    & (1)                     & 5.7 $\pm$ 1.4      & 25000                       & 2.4 (0.4, 9.1)              \\
    LCC           & 558  (2363)        & 0.8 $\pm$ 0.2              & 50-300                  & (2)                     & 4.4 $\pm$ 0.9      & --                          & --                          \\
    Upper-Sco     & 3320 (5203)        & 0.4 $\pm$ 0.1              & 11000                   & App.~\ref{saap:mass}    & 4.1 $\pm$ 1.2      & 70000                       & 3.2 (0.1, 9.1)              \\
    \midrule
    Upper-Sco~I   & 1115 (2998)        & 0.7 $\pm$ 0.2              &                         &                         &                    &                             &                             \\
    Upper-Sco~II  & 1670               & 0.5 $\pm$ 0.1              &                         &                         & 3.3 $\pm$ 0.5      &                             &                             \\
    \bottomrule 
    \end{tabular}
    \tablefoot{In column 2, we list the number of stars formed along each cluster chain excluding the clusters older than \SI{15}{\mega\year}, with values in parenthesis indicating the count when all clusters are included. The average acceleration and its uncertainty in column 3 are computed as the median and median absolute deviation, respectively, of the posterior probability distribution from the linear regression. The residual gas mass at the end of each cluster chain is given in column 4, with their corresponding references in column 5. In Column 6, we list the radial velocity measured for the CO gas \citep{Dame_01_ApJ} within the residual gas structures of the cluster chains (see Appendix~\ref{saap:rv} for details). Columns 7 and 8 lists the current cloud momentum (see Sect.~\ref{sub:momentum} for details) and the median and 2$\sigma$-interval of the number of SNe needed to account for this momentum, respectively. In the case of the LCC chain, we refrain from a momentum estimation, see Sect.~\ref{sub:momentum}.}
    \tablebib{(1) \cite{Alves_14_AA, Posch_23_AA}, (2) \cite{Boulanger_98_AA}}
    \end{table*}

    For an accurate kinematic analysis, compiling a large sample of radial velocities is essential, as well as limiting this selection to stars with precise velocity measurements.
    We combined radial velocity data from the Sloan Digital Sky Survey (SSDS) APOGEE-2 DR17 \citep{Apogee2_22_ApJ} and GALAH DR3 \citep{Galah_21_MNRAS}, with radial velocity data from \textit{Gaia} DR3 \citep{Gaia_23_AA, Katz_23_AA}, when available. For the APOGEE-2 radial velocity error, we added the nominal uncertainty (\texttt{VERR}) in quadrature to the scatter-value for stars with more than one visit (\texttt{VSCATTER}). For stars with multiple radial velocity measurements within these three surveys, we selected the radial velocity with the smallest measurement uncertainty.
    
    We selected a subsample of stars based on their radial velocities and corresponding uncertainties to exclude poor measurements that would impede a precise kinematic analysis.
    The kinematic sample was restricted to stars with radial velocity errors smaller than \SI{1}{\km\per\s}. Within this sample, 97\% of radial velocity measurements in Sco-Cen fall between \num{-50} and \SI{50}{\km\per\s}, and we excluded stars with radial velocity measurements outside this range, removing extreme outliers from the sample.
    Without them, the standard deviation of the radial velocity distribution for each cluster is well-constrained, with $\sigma <$ \SI{12}{\km\per\s} and an average standard deviation of around \SI{6}{\km\per\s}.
    As the analysis is focused on high-confidence cluster members, we further excluded stars with radial velocities exceeding the one-sigma threshold from the remaining radial velocity distribution of each cluster. This ensures a narrow velocity distribution for all clusters, excluding outliers. Clusters with fewer than ten measurements exhibit minimal variation in their radial velocities, all within a couple of \si{\km\per\s} of each other, making further filtering unnecessary.
    
    We converted the observed coordinates and velocities of this data set to heliocentric Galactic Cartesian coordinates and velocities (\textit{X,~Y,~Z,~U,~V,~W}) using the \texttt{astropy} Python library \citep{Astropy_22_ApJ}.


\section{Results}\label{sec:results}

    This section describes the main characteristics of cluster chains: coherent sequences in position, age, speed, and mass. In addition, we present the residual gas structures at the end of all three cluster chains and estimate the cloud momentum.
    

    \subsection{Linear spatio-temporal alignment}\label{sub:spatiotemp}
    The spatio-temporal alignment of the chains of clusters can be seen in Fig.~\ref{fig:sigma-chains}, with older clusters indicated in red and younger clusters in blue. In the CrA and LCC chains (left and middle panels), we observe a distinct and well-ordered age gradient along each cluster chain, from the older progenitor clusters at the center to younger clusters at the outskirts of Sco-Cen. The Upper-Sco chain also shares the basic structure in which older clusters are in the inner regions of Sco-Cen and younger clusters extend from the center outwards.
    However, the detailed morphology of this cluster chain is more complicated, and several clusters overlap spatially. We tentatively divided the Upper-Sco chain into two subchains that differ in their kinematic and mass properties, and describe this procedure in the following sections. We discuss the peculiarities of the Upper-Sco chain in Sect.~\ref{ssub:USc}.
    
    \subsection{Cluster acceleration}\label{sub:speed}
\begin{figure}
        \centering
        \includegraphics[width=\columnwidth]{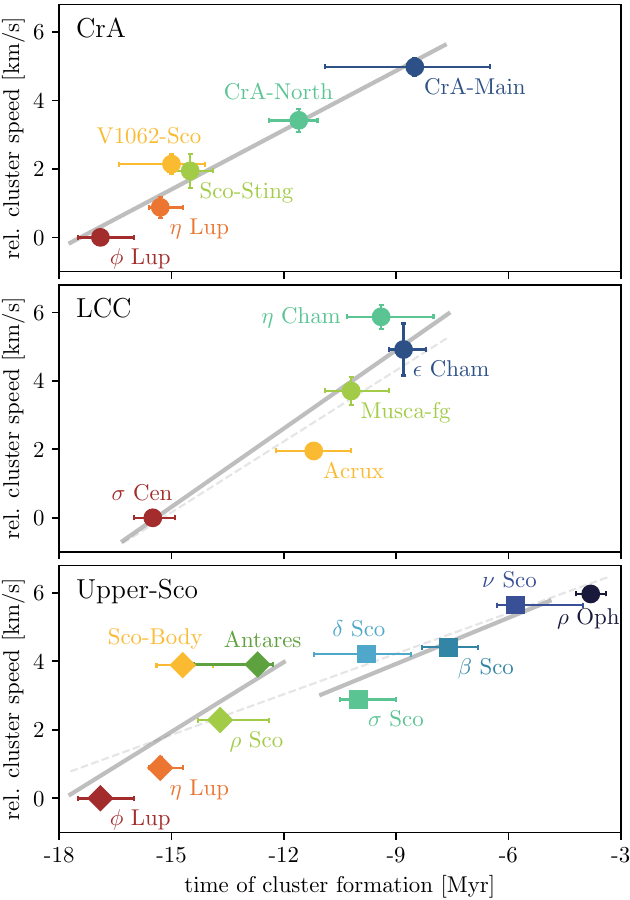}
        \caption{Cluster speed relative to the motion of the progenitor clusters of each chain, plotted against the time of cluster formation.
        We estimate the average acceleration along each cluster chain with linear regression curves (gray lines), including two alternative fits to the data (light-gray dashed lines). See Sect.~\ref{sub:speed} for details. We list the gradients in Table~\ref{tab:chains} and all fitting parameters in Table~\ref{tab:param}. In the bottom panel, we indicate the two subchains Upper-Sco~I and Upper-Sco~II with diamonds and squares, respectively.}
        \label{fig:speed_time}
    \end{figure}

    An important characteristic of cluster chains is that they exhibit a correlation between cluster speed and age, in which the youngest clusters, furthest from the center of Sco-Cen, are moving away at higher velocities.
    This relation is shown in Fig.~\ref{fig:speed_time}, where we plot the relative speed of each cluster against the time of cluster formation.
    The relative cluster speed was computed as the vector modulus of the relative cluster velocity with respect to that of the progenitor clusters (PrC) within each chain ($|(\text{UVW}) - (\text{UVW})_{\text{PrC}}|$).
    This computation was done statistically and we refer to Appendix~\ref{app:prop_tables} for details on the method. Fig.~\ref{fig:speed_time} shows the median of the relative speed distribution, with error bars representing the median absolute deviation.
    The average motion of clusters in Sco-Cen older than \SI{15}{\mega\year} (SC-15) was used as an alternative reference frame and we list cluster speed measured relative to both reference frames in Table~\ref{tab:relSpeed}.    
    
    Figure~\ref{fig:speed_time} shows that the cluster speed relative to the progenitor cluster's motion increases with decreasing cluster age. This correlation indicates an acceleration along each cluster chain, and the youngest clusters move away the fastest from the origin of each chain. 
    We applied a Bayesian linear regression to the data points in Fig.~\ref{fig:speed_time} to estimate this acceleration, along with an associated uncertainty (detailed in Appendix~\ref{saap:acceleration}). The slope of the regression line gives the average acceleration for each cluster chain, as the acceleration of each cluster chain likely varied over time. The resulting acceleration is listed in Table~\ref{tab:chains} and the full set of regression parameters is given in Table~\ref{tab:param}.
    
    The average acceleration in Upper-Sco is slightly lower (light-gray dashed line) compared to that of the CrA and LCC chains but remains compatible with both.
    Based on the speed and mass relations (Sect.~\ref{sub:mass}), we visually identified two consecutive sequences with varying slopes. We tentatively divided the Upper-Sco chain into two subchains: Upper-Sco~I, which is older (\SIrange{12}{18}{\mega\year}, diamonds) and has a higher acceleration, and Upper-Sco~II, which is younger (\SIrange{5}{11}{\mega\year}, squares) and has a lower acceleration. The average acceleration of the two subchains in Upper-Sco are comparable to those of the CrA and LCC chains. We plot the fits to both subchains in Fig.~\ref{fig:speed_time}.
    We also find that $\eta$~Cham deviates from the general relative speed-age relation in LCC. Excluding this cluster from the fit reduces the acceleration of the LCC chain to about
    \SI{0.7}{\km\per\s\per\mega\year} (light-gray dashed line), which is still slightly above the average for all chains of clusters.

    For comparison, we computed the acceleration along the chains with the velocities measured relative to the average motion of SC-15 (see Appendix~\ref{app:prop_tables}). We observe that only the acceleration along the LCC chain slightly changed by \SI{0.1}{\km\per\s\per\mega\year}. This can be attributed to the velocity of $\sigma$~Cen deviating from to the average motion of SC-15, whereas the velocity of $\phi$~Lup aligns more closely with it.
    
    The average acceleration of all chains, independent of reference frame, is about
    \SI{0.6}{\km\per\s\per\mega\year} leading to an increase in cluster speed of about
    \SI{6}{\km\per\s} over the \SI{10}{\mega\year} during which the cluster chains were formed.
    The acceleration along cluster chains contributes to the general increase in cluster speed versus age as found by Gro\ss schedl et al in prep. for the entire Sco-Cen association.

    \begin{figure}
        \centering
        \includegraphics[width=\columnwidth]{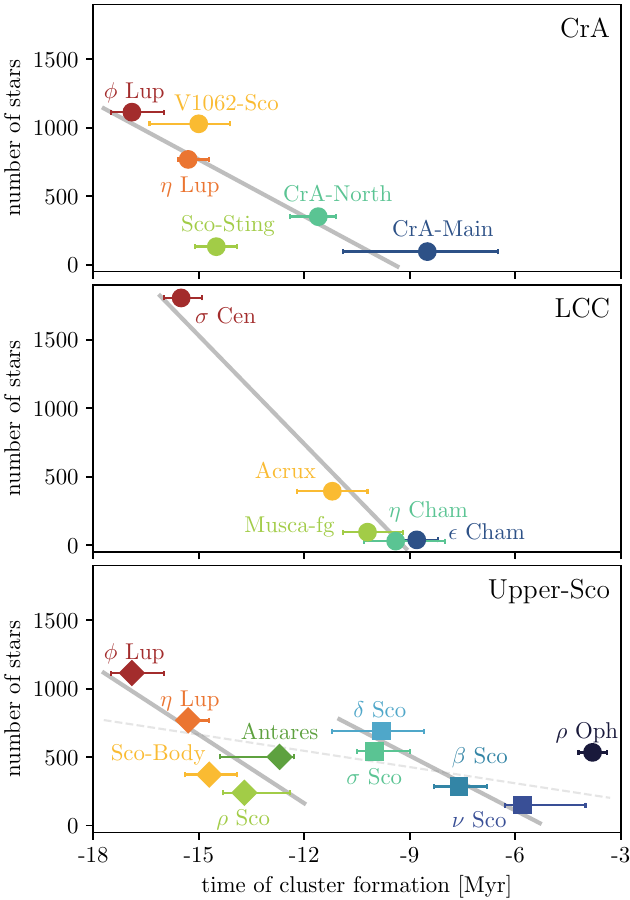}
        \caption{The number of stars per cluster versus the time of cluster formation. For Upper-Sco, we indicate the two subchains Upper-Sco~I and Upper-Sco~II with diamonds and squares, respectively. Linear regression curves indicate the average decrease in cluster mass for each chain, with fitting parameters being listed in Table~\ref{tab:param}}.
        \label{fig:mass_time}
    \end{figure}

    \subsection{Decline in cluster mass}\label{sub:mass}
    The correlation between cluster mass and age is another important characteristic of cluster chains. In Fig.~\ref{fig:mass_time} we show the number of cluster members plotted against the time of cluster formation.
    We use the number of cluster members as an approximation for the cluster mass to avoid uncertainties associated with computing individual stellar masses through isochrone fitting.
    The total number of cluster members can then be approximated into a cluster mass by multiplying it with an average stellar mass factor of about \SI{0.42}{\solarmass} based on the initial mass function from \cite{Kroupa_01_MNRAS}. We adopted a Poissonian error, $\sqrt X$, as uncertainty estimation for the number of stars.

    Figure~\ref{fig:mass_time} shows a distinct decrease in cluster mass with age for each cluster chain.
    To quantify this average decrease in star formation rate, we fitted linear regression curves to the data. The fitting parameters for each cluster chain are listed in Table~\ref{tab:param}. We find an average decrease in cluster size of about \num{-180} stars Myr$^{-1}$, and multiplying this value by an average stellar mass of about \SI{0.42}{\solarmass}, this translates to a decrease in star formation rate of approximately \SI{-80}{\solarmass\per\mega\year}.
    
    We observe a larger dispersion for the cluster mass trend within the Upper-Sco chain of clusters. However, this scatter is significantly reduced when we split the mass-age relation into the two subchains previously defined based on cluster acceleration. Upper-Sco~I shows a mass decrease from $\phi$~Lup to $\rho$~Sco and Antares (about \SI{-70}{\solarmass\per\mega\year}), followed by a second mass relation that begins with higher masses in $\delta$~Sco and $\sigma$~Sco and decreases until $\nu$~Sco (about \SI{-60}{\solarmass\per\mega\year}). $\rho$~Oph seems to be the inception of another mass sequence. This supports the previously established existence of two subchains in Upper-Sco, possibly with another currently forming.

    \subsection{Residual gas structures}\label{sub:cloud}
\begin{figure*}
        \centering
        \includegraphics[width=\textwidth]{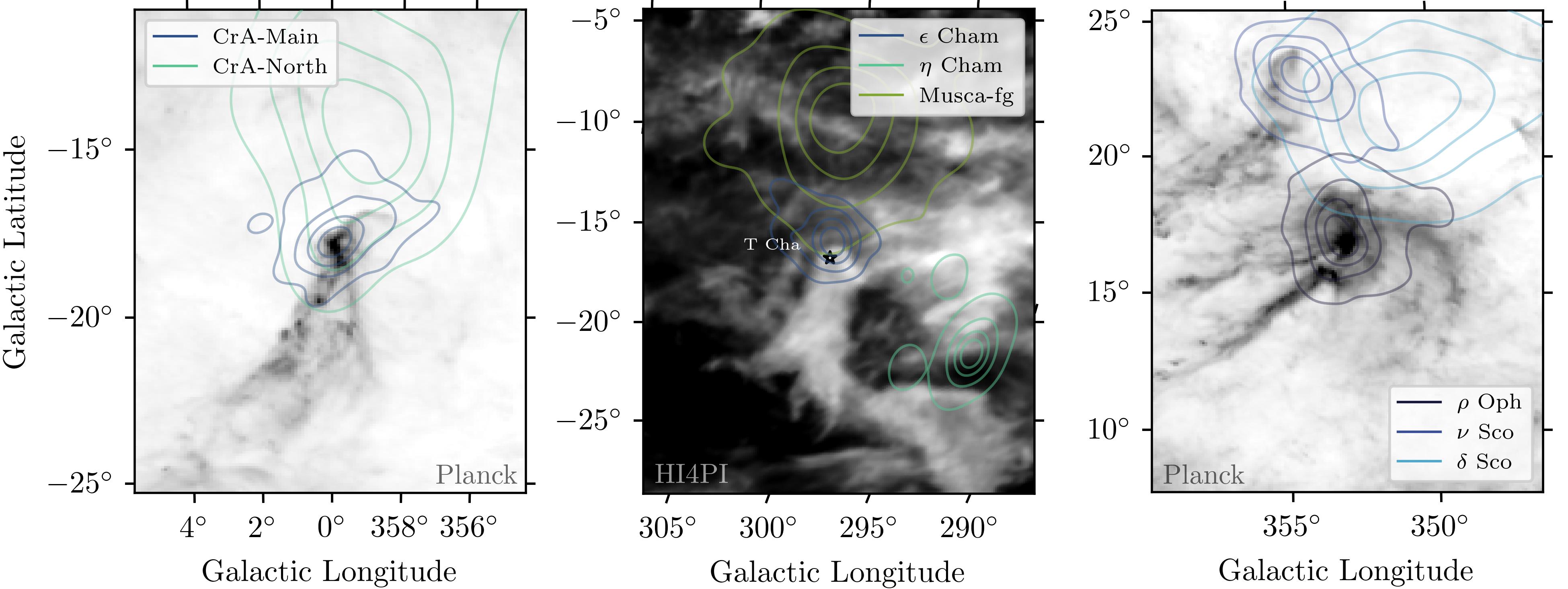}
        \caption{Planck \SI{857}{\giga\hertz} images \citep{Planck_14_AA, Planck_20_AA} of the cometary-shaped residual gas structures at the end of the CrA and Upper-Sco cluster chains in the left and right panels, respectively. In the center panel, the residual molecular cloud at the end of the LCC chain, the Blue Cloud \citep{Boulanger_98_AA, Nehme_08_AA}, is shown in HI gas emission using the HI4PI data at about \SI{5}{\km\per\s} \citep{Hi4Pi_16_AA}. Image dimensions are adapted to show the same physical \num{30}x\SI{40}{\parsec} ratio at varying distances, using the distance to the youngest associated clusters CrA~Main (\SI{155}{\parsec}), $\epsilon$~Cham (\SI{102}{\parsec}), and $\rho$~Oph (\SI{139}{\parsec}). The stellar density distributions of the associated clusters of each chain are indicated with contour lines (5\%, 25\%, 50\%, and 75\%) on top.}
        \label{fig:cloud}
    \end{figure*}

    Figure~\ref{fig:cloud} shows the residual gas structures associated with the youngest cluster of each cluster chain. They share the same wind-blown morphology, with dense heads and dispersed tails, and host at least one YSO within the cloud's head.

    The CrA molecular cloud and the Ophiuchus molecular cloud are the two remaining clouds associated with CrA Main and $\rho$~Oph, respectively, and are shown in Fig.~\ref{fig:cloud} using Planck \SI{857}{\giga\hertz} data. Both remnant clouds are currently forming a new star cluster, the Coronet and $\rho$~Oph/L1688 clusters, respectively \citep[e.g., Coronet cluster][]{Neuhauser_Forbich_08, Sandell_21_ApJ, Sabatini_24_AA} \citep[e.g., $\rho$~Oph/L1688 cluster][Alves et al. in prep.]{Casanova_95_ApJ, Lombardi_08b_AA, Ladjelate_20_AA, Grasser_21_AA}. The molecular cloud Barnard 40 (B40), located at (l, b)~$\approx$~(\num{356}, \num{22})~\si{\degree} in the right panel of Fig.~\ref{fig:cloud}, is likely the remnant structure of the $\nu$~Sco cluster, with the B-type star $\nu$~Sco illuminating the surrounding gas.

    Contrary to the CrA and Ophiuchus molecular clouds, the residual gas structure at the end of the LCC chain is not clearly distinguishable from the background clouds in the Planck map. In a study of the Chameleon molecular clouds, \cite{Boulanger_98_AA} discussed a faint foreground gas cloud with a velocity different from that of the Chameleon clouds in the background. \cite{Nehme_08_AA} later conducted a detailed study of this diffuse structure, naming it the Blue Cloud, and suggesting a connection to LCC. The Blue Cloud is located at the end of the LCC chain and hosts the embedded YSO T~Cha \citep{Franchini_92_AA, Covino_97_AA, Mizuno_01_PASJ, Schisano_09_AA, Hendler_18_MNRAS}, which is commonly associated with $\epsilon$~Cham \citep{LopezMartin_13_AA, Murphy_13_MNRAS, Dickson_21_AJ, Varga_24_arXiv}. 
    We confirm the radial velocity difference between the Blue Cloud and the Chameleon molecular clouds in the background using CO data from \cite{Dame_01_ApJ} and HI data from the HI4PI survey \citep{Hi4Pi_16_AA}. While the Blue Cloud is visible in a narrow velocity range between \SIrange{4}{5}{\km\per\s} in both CO and HI data, the Chameleon clouds show wider ranges in CO data: \SIrange{0}{4}{\km\per\s} for Cham~II and Cham~III, and \SIrange{3}{6}{\km\per\s} for Cham~I. In Fig.~\ref{fig:cloud}, the Blue Cloud is shown in HI emission, with Cham~I to the right. 
    Additionally, \citetalias{Ratzenboeck_23a_AA} further suggests that the YSO T~Cha is a member of $\epsilon$~Cham similar to previous studies. Combined, these observations strongly suggest that the Blue Cloud is the residual gas structure of the LCC chain.

    
   
    The radial velocities of the youngest clusters of each cluster chain align with those of their associated molecular clouds. The radial velocities of CrA~Main, $\epsilon$~Cham, $\nu$~Sco, and $\rho$~Oph match those of the CrA molecular cloud, the Blue Cloud, the B40 molecular cloud, and the Ophiuchus molecular cloud within the uncertainties, respectively.
    Details are outlined in Appendix~\ref{saap:rv}. The consistency between cloud and cluster motion supports the assumed connection between young clusters and their parent molecular clouds and complements similar observations of other regions \citep[e.g.,][]{Tobin_09_ApJ, Hacar_16_AA, Grossschedl_21_AA}.
    
    \subsection{Momentum analysis of cluster chains}\label{sub:momentum}
    Similarly to \cite{Grossschedl_21_AA} and \citetalias{Posch_23_AA}, we estimated the momentum that the Ophiuchus cloud has today, based on its current mass and the average velocity of its associated young stars. Additionally, we estimated the number of SNe needed to account for this momentum, as the impact of SN explosions likely dominated stellar feedback processes. While \citetalias{Posch_23_AA} considered stellar winds from B-type stars in their analysis, they found the contribution to be negligible. We refer to Appendix~\ref{app:momentum} for more details on the probabilistic computation of each estimate. The results are summarized in Table~\ref{tab:chains}.

    We estimated a mass of about \SI{11000}{\solarmass} for the Ophiuchus molecular cloud based on the 3D dust maps from \cite{Leike_20_AA} and \cite{Edenhofer_24a_AA}. We determined a relative velocity of $\sim$~\SI{6}{\km\per\s} for $\rho$~Oph with respect to both reference frames, $\phi$~Lup and SC-15, and adopted this as the velocity of the molecular cloud.
    We obtained a median momentum of around \SI{70000}{\km\per\s\solarmass} for the Ophiuchus molecular cloud, multiplying the mass by the velocity. Assuming a cloud size of \SIrange{8}{10}{\parsec} and a distance to the source of feedback of \SI{15\pm5}{\parsec}, this momentum estimate can be attributed to a median number of approximately three SN explosions, with 95\% of the distribution lying between one and nine SNe. 

    For the residual cloud at the end of the LCC chain, we considered mass measurements reported by \cite{Boulanger_98_AA} that include the cloud's head and tail. They measured the molecular hydrogen mass ($\sim$~\SI{300}{\solarmass}) based on \SI{100}{\micro\meter} radiation and the atomic hydrogen mass from the same area ($\sim$~\SI{50}{\solarmass}). The molecular hydrogen measurement is highly uncertain because the Chameleon molecular clouds dominate the background.
    We cannot estimate the momentum of the Blue Cloud because most of its mass is dispersed against the background, making it difficult to obtain a reliable mass estimate.
    However, based on momentum conservation, we can infer the mass of the Blue Cloud by assuming that the cluster chains experienced similar momentum injection. Given the similarity between the LCC and CrA chains, we adopted the momentum of the CrA molecular cloud (\SI{25000}{\km\per\s\per\solarmass}) as an upper limit for the momentum of the Blue Cloud. Using the velocity of $\epsilon$~Cham relative to that of $\sigma$~Cen and SC-15 as the current cloud velocity ($\sim$~\SIrange{5}{7}{\km\per\s}), we estimated a potential cloud mass of \SIrange{3500}{5000}{\solarmass}. This estimate is significantly larger than the maximum observed mass by \cite{Boulanger_98_AA}, suggesting that up to 95\% of the gas mass was dispersed below a threshold where cloud boundaries become difficult to define.

\section{Discussion}\label{sec:discussion}

    The CrA cluster chain was the first coherent structure of sequential cluster formation with a detailed characterization of its kinematics and origin \citepalias{Posch_23_AA}.      
    Here, we investigate two additional cluster chains, LCC and Upper-Sco, which show similar characteristics to the CrA chain. Our analysis reveals well-defined patterns: 1) an approximate linear spatial distribution, 2) a sequential age gradient, 3) a speed gradient increasing with time, and 4) a mass gradient that decreases from older to younger clusters. Collectively, these properties seem to define a new type of stellar structure that is an important component of OB associations in an evolutionary stage similar to that of Sco-Cen.
    We propose that each of the three chains — CrA, LCC, and Upper-Sco — represents different versions of the same phenomenon: triggered star formation along residual gas structures driven by stellar feedback, with differences attributed to the variation in properties of the progenitor cloud and feedback forces.  
    
    In addition to the three cluster chains examined in this study, \cite{MiretRoig_24_AA} have recently identified the well-known TW Hydra association as another cluster chain in Sco-Cen. This new chain shares the same spatio-temporal and kinematic properties discussed in this work. Recently, \cite{Pang_21_ApJ} and \cite{Kerr_21_ApJ, Kerr_24_arXiv} identified large-scale spatio-temporal cluster sequences in the Gamma Vel, Sco-Cen, and the Cepheus-Hercules complexes, reminiscent of the results presented here, although at limited resolution for the more distant regions. It is then safe to speculate that the formation of cluster chains is a common mode of star formation in stellar associations that contain massive stars, which are capable of sustaining feedback over timescales of \SIrange{5}{10}{\mega\year}. Finally, given the well-defined temporal structure of cluster chains, they appear to be ideal laboratories for studying chemical enrichment in star-forming regions \citep[e.g.,][]{Spina2017-xl, Kos_21_MNRAS}.

    \subsection{Formation of cluster chains}\label{sub:formation}
    The occurrence of multiple similar stellar structures, exhibiting the same physical properties, suggests a common formation mechanism. In this section, we expand and generalize the formation scenario of the CrA chain described in \citetalias{Posch_23_AA}, to explain the physical properties of all three cluster chains in Sco-Cen. This scenario is largely based on previous models for triggered star formation \citep{Elmegreen_Lada_77_ApJ, Whitworth_94_AA}. However, we focus here on describing a specific mode in which feedback creates cometary-shaped clouds, or pillars, compresses, and pushes them away, creating cluster sequences over time. This process proceeds from the center of an association towards its periphery. 
        
    In this scenario, we start with the residual molecular gas surrounding the clusters formed in the main star formation event of an OB association. Stellar feedback from these massive clusters, including stellar winds (ram pressure), radiation pressure, photoionization, and SN explosions, create an ISM flow \citep[e.g.,][]{Redfield_Linsky_08_ApJ, Nehme_08_AA, Piecka_24_AA}, sustained over several million years, which acts as an anisotropic pressure on the remaining molecular gas clouds. In this paper, we will not distinguish which type of feedback plays the dominant role in Sco-Cen.
    These forces i) push the gas away from the source of feedback and ii) compress the part of the cloud that is closest to the source of feedback. As a result, the gas is shaped into a cometary cloud, with the dense head shielding a collapsing core from disruptive ultraviolet radiation and thus enabling star formation.
    This process is similar to the formation of pillars found in HII regions \citep[e.g.,][]{Hosokawa_Inutsuka_06_ApJ, Dale_07_MNRAS, Bisbas_11_ApJ, Elmegreen_11_EAS, Rogers_Pittard_14_MNRAS, Hartigan_15_AJ}, although there are few signs of ionized gas surrounding the remnant gas clouds in Sco-Cen \citep[e.g.,][Alves et al. in prep.]{Nehme_08_AA}. This may suggest that different processes, such as winds and supernovae (SNe), rather than photoionization, are currently the primary processes in Sco-Cen.

    Over time, the cloud is continuously affected by this feedback flow, unlike the stars it formed.
    This feedback leads to the sequential formation of clusters until the gas is either fully depleted or dispersed. In this sequence, younger clusters exhibit lower masses due to the diminishing gas reservoir available for star formation. This observation agrees with the correlation between the cloud mass and the star formation rate \citep{Schmidt_59_ApJ, Lada_12_ApJ, Roccatagliata_13_AA, Zamora_Vazquez_14_ApJ}.
    Once a cluster emerges from its progenitor cloud, its motion is no longer influenced by the feedback flow, thus reflecting the cloud's velocity at the time of the cluster's formation. Consequently, younger clusters exhibit higher velocities away from the feedback source. This scenario naturally reproduces the observed sequential spatio-temporal nature of cluster chains in Sco-Cen, and also explains the speed and mass gradients.  

    Additional evidence for Sco-Cen's ability to sustain a long-lived feedback flow comes from observations tracing recently ejected runaway stars, likely arising from supernova explosions in Sco-Cen. \cite{Neuhauser_20_MNRAS} traced several post-SN candidate neutron stars to Upper-Sco and $\sigma$~Cen/LCC, suggesting they exploded around \SI{2}{\mega\year} ago. In Upper-Sco, \cite{Briceno_23_MNRAS} predicted four SN explosions within the past \SI{10}{\mega\year}, including the one that caused the runaway star $\zeta$~Oph. Previous studies agree on a number of around 10--20 SN explosions in Sco-Cen, creating the Local Bubble \citep[e.g.,][]{Fuchs_06_MNRAS, Krause_18_AA, Zucker_22_Natur, Swiggum_24_Nature}. In a follow-up paper, we will investigate the relative roles of stellar feedback and the Galactic potential \citep[e.g.,][]{Konietzka_24_Nature}, and quantify their impact on the observed acceleration in cluster chains.

    \subsection{Variations within cluster chains}\label{sub:variation}
    Although all cluster chains were likely formed via the same mechanism, we observe some differences in their properties, such as acceleration rates, the number of stars formed and the decrease in star formation rate, current estimated momentum, and their potential for continuing star formation. These variations may be attributed to different environmental conditions, such as the initial cloud mass, the distance from the sources of feedback, and the amount of feedback experienced by each cloud. 
    
    The CrA and LCC chains have similar ages and number of stars, but the gas reservoir at the end of the CrA chain currently retains on average approximately 50 times more mass than the Blue Cloud at the end of LCC \citep{Boulanger_98_AA, Alves_14_AA, Posch_23_AA}. Additionally, while the CrA molecular cloud continues to form stars \citep[e.g.,][]{Taylor_Storey_84_MNRAS, Neuhauser_Forbich_08, Tachihara_24_ApJ, Sabatini_24_AA}, only 10\% of the residual gas associated with the LCC chain is currently molecular \citep{Boulanger_98_AA}, indicating that star formation is effectively finished. T Cha, still embedded within the cloud's head, might be the final star formed in the LCC chain.
    The Upper-Sco chain, is the most massive of the three cluster chains, having formed at least 3~300 stars after the main star formation event in Sco-Cen \SI{15}{\mega\year} ago. It has the largest remaining gas reservoir of the three chains, and substantial star formation is currently taking place at the head of the Ophiuchus clouds (L1688/ $\rho$~Oph core, L1689, L1709, e.g., Alves et al. in prep.).

    We note that the $\eta$~Cham cluster, part of the LCC chain, deviates from the general relations in position and velocity. It has a slightly older age than $\epsilon$~Cham (\citealt{Lawson_09_MNRAS}, \citealt{Murphy_13_MNRAS}, \citetalias{Ratzenboeck_23b_AA}) while being the furthest away and moving away the fastest from Sco-Cen. The gas cloud that formed this cluster might have been a different, smaller cloud in the complex, or differently affected by the feedback.

    We observe little variation in the CrA chain, except for the positional offset of V1062-Sco,
    and the deviation of Sco-Sting from the otherwise well-constrained mass relation. 

        \subsubsection*{Upper-Sco likely consists of two or more cluster chains}\label{ssub:USc}        
        
        Upper-Sco is the most complex cluster chain, containing six times more stars than the CrA and LCC chains, with several clusters overlapping in their 3D positions. We distinguished between two formation episodes by analyzing this cluster chain's speed and mass relations. Upper-Sco~I (\SIrange{12}{18}{\mega\year}) shows a fast acceleration, likely driven by a large amount of stellar feedback from the massive clusters at the center of Sco-Cen, that formed during the main star-forming event around \SI{15}{\mega\year} ago. Starting star formation several million years later, Upper-Sco~II (\SIrange{5}{11}{\mega\year}) shows lower acceleration, possibly due to experiencing less stellar feedback. $\rho$~Oph potentially begins a new subchain, exhibiting continued acceleration and a substantial gas reservoir still available for future star formation.
       
        We now discuss a possible scenario to explain the dual speed and mass relationships in the Upper-Sco chain. Feedback from the main star formation event in Sco-Cen triggered star formation in the lower-density gas surrounding the massive clusters $\eta$~Lup and $\phi$~Lup. This gas was rapidly accelerated away from the feedback source, sequentially forming the Sco-Body, $\rho$~Sco, and Antares clusters. Meanwhile, the expanding feedback bubble accumulated molecular material or encountered a cloud in the Sco-Cen complex, creating the massive progenitor molecular cloud of Upper-Sco~II. After becoming unstable, it collapsed and formed $\sigma$~Sco and $\delta$~Sco, which collectively host more than 1~000 stars, enabling the formation of another cluster chain. The massive stars in these clusters contributed — and some still contribute — to the feedback flow that accelerated the remaining molecular material, leading to the formation of $\beta$~Sco and $\nu$~Sco. The molecular cloud B40 might be the residual gas of Upper-Sco~II.       
        
        Investigating the motion of clusters in Upper-Sco, we found that the Ophiuchus cloud,
        currently forming the closest embedded stellar cluster to Earth, L1688,
        was within \SI{3\pm1}{\parsec} of the $\delta$~Sco cluster approximately \SI{6}{\mega\year} ago. Details on the traceback are given in Appendix~\ref{app:rhoOph}. This confirms the previous estimate from \cite{MiretRoig_22_AA} who found an even closer proximity between $\delta$ and $\rho$~Oph, of about \SI{1}{\parsec} around \SI{5}{\mega\year} ago. Furthermore, we find that $\rho$~Oph was within \SI{9\pm1}{\parsec} of the $\beta$~Sco cluster about \SI{6}{\mega\year} ago, and within \SI{12\pm1}{\parsec} of the $\sigma$~Sco cluster about \SI{4}{\mega\year} ago, approximately when the $\rho$~Oph cluster started to form. Both $\sigma$~Sco and $\delta$~Sco currently host massive stars: $\delta$~Sco is a B0.2IVe star with an apparent magnitude of V$=$\SI{2.32}{\mag}, and $\sigma$~Sco is a B1III+B1V star with V$=$\SI{2.89}{\mag}, likely influencing the gas structure around $\rho$~Oph (e.g., Alves et al. in prep.).
        This proximity to past and present feedback sources further supports the proposed formation scenario of the Upper-Sco~II subchain.
        $\rho$~Oph may be the first cluster in a new chain of clusters that is forming from the gas currently visible as L1688, L1689, and L1709.
        
        Recognizing Upper-Sco as a cluster chain aligns with previous work on the formation of this star-forming region. It has been proposed that the formation of Upper-Sco, as one group, was triggered by the central clusters in UCL \citep[e.g.,][]{Preibisch_99_AJ, Fernandez_08_AA, Krause_18_AA}. The discovery of multiple clusters within Upper-Sco has clarified its star formation history, revealing a sequential process lasting around \SI{10}{\mega\year} that created the region's observed substructure \citep[e.g.,][]{Squicciarini_21_MNRAS, MiretRoig_22_AA, Briceno_23_MNRAS}.
        We find that the conventional Upper-Sco clusters, which we identify as the second subchain Upper-Sco~II, are part of a larger structure that connects them to the massive central clusters of Sco-Cen, implying sequential star formation for a period of approximately \SI{15}{\mega\year}.
        
        In its youngest two clusters, the Upper-Sco chain deviates from a linear spatial distribution, showing a decrease in distance from the Galactic plane rather than continuing to extend away from it. $\nu$~Sco and $\rho$~Oph show a strong velocity change towards the Galactic plane, which suggests a more complex formation scenario. This could be due to the Galactic potential pulling the gas reservoir back toward the Galactic plane or interactions between the molecular material around \SIrange{5}{6}{\mega\year} ago. These observations indicate the need for further detailed studies.
        
    \subsection{What fraction of stars in Sco-Cen formed via triggered star formation?}\label{sub:triggered}

    The fraction of stars formed through triggered star formation is a classical and still unresolved question in star formation research. We are able to address the question in this paper, as our findings suggest that cluster chains are the product of triggered star formation.
    Excluding the massive clusters at the beginning of each cluster chain older than \SI{15}{\mega\year} (V1062-Sco, $\eta$ and $\phi$~Lup, and $\sigma$~Cen), we find that about 34\% (4457 stars) of the total stellar population of Sco-Cen\footnote{We exclude the three clusters Oph-Southeast, Oph-NorthFar, and Norma-North since they are likely unrelated to Sco-Cen \citepalias{Ratzenboeck_23b_AA}.} (12972 stars) is formed in cluster chains as a result of stellar feedback.
    However, multiple smaller clusters such as the Pipe (B59), Chamaeleon, and Lupus 1-4 clusters, are not part of cluster chains studied here and likely originate from the vicinity of older massive clusters in Sco-Cen as well \citep[e.g.,][Alves et al. in prep.]{Edenhofer_24b_AA, Bobylev_24_arxiv}. If we include these regions, the fraction of triggered star formation in Sco-Cen increases to 39\%. Our results indicate that a significant fraction of stars are formed concurrently during the destruction of the remaining cloud structures by stellar feedback.

    This percentage is in good agreement with the findings of \cite{Snider_09_ApJ} using \textit{Spitzer} data. They estimated that 25-50\% of stars form via triggered star formation, based on YSO overdensities associated with compressed gas in HII regions. Our results exceed those found by \cite{Getman_12_MNRAS} in the Elephant trunk nebula in which they estimated 14-25\% of stars formed via triggered star formation. \cite{Thompson_12_MNRAS} and \cite{Kendrew_12_ApJ} observed the percentage of massive stars formed through triggered star formation and found fractions of 14-30\% and 20\%, respectively. However, their results cannot be directly compared to ours, which include low-mass stars.

    We have identified two modes of star-formation in Sco-Cen: 1) the bulk of the stellar association including the progenitor clusters within each chain, formed about \SIrange{15}{18}{\mega\year} ago, and extending from it, 2) the chains of clusters characterized in this work, each with distinct kinematic signatures \citepalias[see also][]{Posch_23_AA}.
    Recent findings suggest that Sco-Cen is the youngest subgroup within the $\alpha$~Persei family of clusters, which began forming approximately \SI{60}{\mega\year} ago \citep{Swiggum_24_Nature}. This raises the possibility that the bulk of the association formed due to feedback from older members of this cluster family. However, the distinct kinematical properties indicate that if feedback initiated star formation in Sco-Cen, it would be of a different nature from the feedback processes responsible for forming the cluster chains. This is speculative at this point and remains to be tested.


\section{Conclusion}\label{sec:conclusion}

    In this work, we demonstrate that the Sco-Cen OB association produced coherent stellar structures through the propagation of star formation, resulting in chains of clusters. These cluster chains exhibit relationships across multiple dimensions. From older to younger clusters, the clusters increase in speed, decrease in mass, and show a linear spatial sequence from the center of the association to its outskirts. Although each relationship may not be conclusive on its own, their combined presence across all cluster chains strongly supports evidence for feedback-driven star formation in Sco-Cen, resulting in these new stellar structures.    
    
    The main results of this study are:
    \begin{itemize}
    
    \item Using \textit{Gaia} and ancillary archival data, we have characterized three chains of clusters within Sco-Cen, previously identified in \citetalias{Ratzenboeck_23b_AA}: the CrA, LCC, and Upper-Sco chains. The CrA chain was previously characterized in \citetalias{Posch_23_AA}.  
    \item All three cluster chains display distinctive features: 1) linear spatial distributions, 2) speed gradients, 3) sequential age gradients, and 4) mass profiles that decrease from older to younger clusters. Additionally, a cometary-shaped cloud remnant is associated with the youngest cluster of each cluster chain.
    \item The Upper-Sco chain exhibits the most complex substructure. We observe indications of two spatially overlapping subchains that formed sequentially, and the possibility of a third subchain forming from the Ophiuchus molecular cloud. 
    \item Around 40\% of the stars within Sco-Cen are formed as a result of triggered star formation, with 35\% of the population forming in the discussed chains of clusters. 
    \item Our results establish cluster chains as a naturally occurring phenomenon in OB associations, likely created by stellar feedback.   
    
    \end{itemize}

    Finally, Sco-Cen, the nearest OB association to Earth ($\sim$\SIrange{100}{200}{\parsec}), provides an unparalleled chance to investigate star formation at high resolution, both spatially and temporally. The newly identified cluster chains, stellar structures with well-defined ages gradients stretching over \SI{15}{\mega\year}, serve as exceptional laboratories for exploring the intricate processes of chemical enrichment in star-forming regions.

\begin{acknowledgements}
    We thank the anonymous referee for their thoughtful and constructive review. Co-funded by the European Union (ERC, ISM-FLOW, 101055318). Views and opinions expressed are, however, those of the author(s) only and do not necessarily reflect those of the European Union or the European Research Council. Neither the European Union nor the granting authority can be held responsible for them. 
    S.~Ratzenb{\"o}ck acknowledges funding by the Austrian Research Promotion Agency (FFG, \url{https://www.ffg.at/}) under project number FO999892674.
    JG acknowledges co-funding from the European Union, the Central Bohemian Region, and the Czech Academy of Sciences, as part of the MERIT fellowship (Project 101081195, MSCA-COFUND Horizon Europe). JG acknowledges the Collaborative Research Center 1601 (SFB 1601) funded by the Deutsche Forschungsgemeinschaft (DFG, German Research Foundation) – 500700252.
    This research has made use of \textit{Python}, \url{https://www.python.org}, of Astropy, a community-developed core \textit{Python} package for Astronomy \citep{astropy_18, Astropy_22_ApJ}, NumPy \citep{numpy_20}, Matplotlib \citep{matplotlib_07}, Galpy \citep{galpy_15}, Scipy \citep{Scipy_20_NMeth}, PyMC \citep{PyMC_16}, and Plotly \citep{plotly}. Further, it has made use of the SIMBAD database operated at CDS, Strasbourg, France (Wenger et al. 2000), of “Aladin sky atlas” developed at CDS, Strasbourg Observatory, France (Bonnarel et al. 2000; Boch \& Fernique 2014), and of TOPCAT, an interactive graphical viewer and editor for tabular data (Taylor 2005).
\end{acknowledgements}
\bibliographystyle{aa.bst}
\bibliography{AA_2024_51312.bib}

\begin{thebibliography}{125}
\expandafter\ifx\csname natexlab\endcsname\relax\def\natexlab#1{#1}\fi

\bibitem[{{Abdurro'uf} {et~al.}(2022){Abdurro'uf}, {Accetta}, {Aerts}, {Silva Aguirre}, {Ahumada}, {Ajgaonkar}, {Filiz Ak}, {Alam}, {Allende Prieto}, {Almeida}, {Anders}, {Anderson}, {Andrews}, {Anguiano}, {Aquino-Ort{\'\i}z}, {Arag{\'o}n-Salamanca}, {Argudo-Fern{\'a}ndez}, {Ata}, {Aubert}, {Avila-Reese}, {Badenes}, {Barb{\'a}}, {Barger}, {Barrera-Ballesteros}, {Beaton}, {Beers}, {Belfiore}, {Bender}, {Bernardi}, {Bershady}, {Beutler}, {Bidin}, {Bird}, {Bizyaev}, {Blanc}, {Blanton}, {Boardman}, {Bolton}, {Boquien}, {Borissova}, {Bovy}, {Brandt}, {Brown}, {Brownstein}, {Brusa}, {Buchner}, {Bundy}, {Burchett}, {Bureau}, {Burgasser}, {Cabang}, {Campbell}, {Cappellari}, {Carlberg}, {Wanderley}, {Carrera}, {Cash}, {Chen}, {Chen}, {Cherinka}, {Chiappini}, {Choi}, {Chojnowski}, {Chung}, {Clerc}, {Cohen}, {Comerford}, {Comparat}, {da Costa}, {Covey}, {Crane}, {Cruz-Gonzalez}, {Culhane}, {Cunha}, {Dai}, {Damke}, {Darling}, {Davidson}, {Davies}, {Dawson}, {De Lee}, {Diamond-Stanic}, {Cano-D{\'\i}az}, {S{\'a}nchez},
  {Donor}, {Duckworth}, {Dwelly}, {Eisenstein}, {Elsworth}, {Emsellem}, {Eracleous}, {Escoffier}, {Fan}, {Farr}, {Feng}, {Fern{\'a}ndez-Trincado}, {Feuillet}, {Filipp}, {Fillingham}, {Frinchaboy}, {Fromenteau}, {Galbany}, {Garc{\'\i}a}, {Garc{\'\i}a-Hern{\'a}ndez}, {Ge}, {Geisler}, {Gelfand}, {G{\'e}ron}, {Gibson}, {Goddy}, {Godoy-Rivera}, {Grabowski}, {Green}, {Greener}, {Grier}, {Griffith}, {Guo}, {Guy}, {Hadjara}, {Harding}, {Hasselquist}, {Hayes}, {Hearty}, {Hern{\'a}ndez}, {Hill}, {Hogg}, {Holtzman}, {Horta}, {Hsieh}, {Hsu}, {Hsu}, {Huber}, {Huertas-Company}, {Hutchinson}, {Hwang}, {Ibarra-Medel}, {Chitham}, {Ilha}, {Imig}, {Jaekle}, {Jayasinghe}, {Ji}, {Johnson}, {Jones}, {J{\"o}nsson}, {Katkov}, {Khalatyan}, {Kinemuchi}, {Kisku}, {Knapen}, {Kneib}, {Kollmeier}, {Kong}, {Kounkel}, {Kreckel}, {Krishnarao}, {Lacerna}, {Lane}, {Langgin}, {Lavender}, {Law}, {Lazarz}, {Leung}, {Leung}, {Lewis}, {Li}, {Li}, {Lian}, {Liang}, {Lin}, {Lin}, {Lin}, {Lintott}, {Long}, {Longa-Pe{\~n}a}, {L{\'o}pez-Cob{\'a}}, {Lu},
  {Lundgren}, {Luo}, {Mackereth}, {de la Macorra}, {Mahadevan}, {Majewski}, {Manchado}, {Mandeville}, {Maraston}, {Margalef-Bentabol}, {Masseron}, {Masters}, {Mathur}, {McDermid}, {Mckay}, {Merloni}, {Merrifield}, {Meszaros}, {Miglio}, {Di Mille}, {Minniti}, {Minsley}, {Monachesi}, {Moon}, {Mosser}, {Mulchaey}, {Muna}, {Mu{\~n}oz}, {Myers}, {Myers}, {Nadathur}, {Nair}, {Nandra}, {Neumann}, {Newman}, {Nidever}, {Nikakhtar}, {Nitschelm}, {O'Connell}, {Garma-Oehmichen}, {Luan Souza de Oliveira}, {Olney}, {Oravetz}, {Ortigoza-Urdaneta}, {Osorio}, {Otter}, {Pace}, {Padilla}, {Pan}, {Pan}, {Parikh}, {Parker}, {Peirani}, {Pe{\~n}a Ram{\'\i}rez}, {Penny}, {Percival}, {Perez-Fournon}, {Pinsonneault}, {Poidevin}, {Poovelil}, {Price-Whelan}, {B{\'a}rbara de Andrade Queiroz}, {Raddick}, {Ray}, {Rembold}, {Riddle}, {Riffel}, {Riffel}, {Rix}, {Robin}, {Rodr{\'\i}guez-Puebla}, {Roman-Lopes}, {Rom{\'a}n-Z{\'u}{\~n}iga}, {Rose}, {Ross}, {Rossi}, {Rubin}, {Salvato}, {S{\'a}nchez}, {S{\'a}nchez-Gallego}, {Sanderson}, {Santana
  Rojas}, {Sarceno}, {Sarmiento}, {Sayres}, {Sazonova}, {Schaefer}, {Schiavon}, {Schlegel}, {Schneider}, {Schultheis}, {Schwope}, {Serenelli}, {Serna}, {Shao}, {Shapiro}, {Sharma}, {Shen}, {Shetrone}, {Shu}, {Simon}, {Skrutskie}, {Smethurst}, {Smith}, {Sobeck}, {Spoo}, {Sprague}, {Stark}, {Stassun}, {Steinmetz}, {Stello}, {Stone-Martinez}, {Storchi-Bergmann}, {Stringfellow}, {Stutz}, {Su}, {Taghizadeh-Popp}, {Talbot}, {Tayar}, {Telles}, {Teske}, {Thakar}, {Theissen}, {Tkachenko}, {Thomas}, {Tojeiro}, {Hernandez Toledo}, {Troup}, {Trump}, {Trussler}, {Turner}, {Tuttle}, {Unda-Sanzana}, {V{\'a}zquez-Mata}, {Valentini}, {Valenzuela}, {Vargas-Gonz{\'a}lez}, {Vargas-Maga{\~n}a}, {Alfaro}, {Villanova}, {Vincenzo}, {Wake}, {Warfield}, {Washington}, {Weaver}, {Weijmans}, {Weinberg}, {Weiss}, {Westfall}, {Wild}, {Wilde}, {Wilson}, {Wilson}, {Wilson}, {Wolf}, {Wood-Vasey}, {Yan}, {Zamora}, {Zasowski}, {Zhang}, {Zhao}, {Zheng}, {Zheng}, \& {Zhu}}]{Apogee2_22_ApJ}
{Abdurro'uf}, {Accetta}, K., {Aerts}, C., {et~al.} 2022, \apjs, 259, 35

\bibitem[{{Alves} {et~al.}(2014){Alves}, {Lombardi}, \& {Lada}}]{Alves_14_AA}
{Alves}, J., {Lombardi}, M., \& {Lada}, C.~J. 2014, \aap, 565, A18

\bibitem[{{Astropy Collaboration} {et~al.}(2022){Astropy Collaboration}, {Price-Whelan}, {Lim}, {Earl}, {Starkman}, {Bradley}, {Shupe}, {Patil}, {Corrales}, {Brasseur}, {N{\"o}the}, {Donath}, {Tollerud}, {Morris}, {Ginsburg}, {Vaher}, {Weaver}, {Tocknell}, {Jamieson}, {van Kerkwijk}, {Robitaille}, {Merry}, {Bachetti}, {G{\"u}nther}, {Aldcroft}, {Alvarado-Montes}, {Archibald}, {B{\'o}di}, {Bapat}, {Barentsen}, {Baz{\'a}n}, {Biswas}, {Boquien}, {Burke}, {Cara}, {Cara}, {Conroy}, {Conseil}, {Craig}, {Cross}, {Cruz}, {D'Eugenio}, {Dencheva}, {Devillepoix}, {Dietrich}, {Eigenbrot}, {Erben}, {Ferreira}, {Foreman-Mackey}, {Fox}, {Freij}, {Garg}, {Geda}, {Glattly}, {Gondhalekar}, {Gordon}, {Grant}, {Greenfield}, {Groener}, {Guest}, {Gurovich}, {Handberg}, {Hart}, {Hatfield-Dodds}, {Homeier}, {Hosseinzadeh}, {Jenness}, {Jones}, {Joseph}, {Kalmbach}, {Karamehmetoglu}, {Ka{\l}uszy{\'n}ski}, {Kelley}, {Kern}, {Kerzendorf}, {Koch}, {Kulumani}, {Lee}, {Ly}, {Ma}, {MacBride}, {Maljaars}, {Muna}, {Murphy}, {Norman},
  {O'Steen}, {Oman}, {Pacifici}, {Pascual}, {Pascual-Granado}, {Patil}, {Perren}, {Pickering}, {Rastogi}, {Roulston}, {Ryan}, {Rykoff}, {Sabater}, {Sakurikar}, {Salgado}, {Sanghi}, {Saunders}, {Savchenko}, {Schwardt}, {Seifert-Eckert}, {Shih}, {Jain}, {Shukla}, {Sick}, {Simpson}, {Singanamalla}, {Singer}, {Singhal}, {Sinha}, {Sip{\H{o}}cz}, {Spitler}, {Stansby}, {Streicher}, {{\v{S}}umak}, {Swinbank}, {Taranu}, {Tewary}, {Tremblay}, {Val-Borro}, {Van Kooten}, {Vasovi{\'c}}, {Verma}, {de Miranda Cardoso}, {Williams}, {Wilson}, {Winkel}, {Wood-Vasey}, {Xue}, {Yoachim}, {Zhang}, {Zonca}, \& {Astropy Project Contributors}}]{Astropy_22_ApJ}
{Astropy Collaboration}, {Price-Whelan}, A.~M., {Lim}, P.~L., {et~al.} 2022, \apj, 935, 167

\bibitem[{{Astropy Collaboration} {et~al.}(2018){Astropy Collaboration}, {Price-Whelan}, {Sip{\H{o}}cz}, {G{\"u}nther}, {Lim}, {Crawford}, {Conseil}, {Shupe}, {Craig}, {Dencheva}, {Ginsburg}, {Vand erPlas}, {Bradley}, {P{\'e}rez-Su{\'a}rez}, {de Val-Borro}, {Aldcroft}, {Cruz}, {Robitaille}, {Tollerud}, {Ardelean}, {Babej}, {Bach}, {Bachetti}, {Bakanov}, {Bamford}, {Barentsen}, {Barmby}, {Baumbach}, {Berry}, {Biscani}, {Boquien}, {Bostroem}, {Bouma}, {Brammer}, {Bray}, {Breytenbach}, {Buddelmeijer}, {Burke}, {Calderone}, {Cano Rodr{\'\i}guez}, {Cara}, {Cardoso}, {Cheedella}, {Copin}, {Corrales}, {Crichton}, {D'Avella}, {Deil}, {Depagne}, {Dietrich}, {Donath}, {Droettboom}, {Earl}, {Erben}, {Fabbro}, {Ferreira}, {Finethy}, {Fox}, {Garrison}, {Gibbons}, {Goldstein}, {Gommers}, {Greco}, {Greenfield}, {Groener}, {Grollier}, {Hagen}, {Hirst}, {Homeier}, {Horton}, {Hosseinzadeh}, {Hu}, {Hunkeler}, {Ivezi{\'c}}, {Jain}, {Jenness}, {Kanarek}, {Kendrew}, {Kern}, {Kerzendorf}, {Khvalko}, {King}, {Kirkby}, {Kulkarni},
  {Kumar}, {Lee}, {Lenz}, {Littlefair}, {Ma}, {Macleod}, {Mastropietro}, {McCully}, {Montagnac}, {Morris}, {Mueller}, {Mumford}, {Muna}, {Murphy}, {Nelson}, {Nguyen}, {Ninan}, {N{\"o}the}, {Ogaz}, {Oh}, {Parejko}, {Parley}, {Pascual}, {Patil}, {Patil}, {Plunkett}, {Prochaska}, {Rastogi}, {Reddy Janga}, {Sabater}, {Sakurikar}, {Seifert}, {Sherbert}, {Sherwood-Taylor}, {Shih}, {Sick}, {Silbiger}, {Singanamalla}, {Singer}, {Sladen}, {Sooley}, {Sornarajah}, {Streicher}, {Teuben}, {Thomas}, {Tremblay}, {Turner}, {Terr{\'o}n}, {van Kerkwijk}, {de la Vega}, {Watkins}, {Weaver}, {Whitmore}, {Woillez}, {Zabalza}, \& {Astropy Contributors}}]{astropy_18}
{Astropy Collaboration}, {Price-Whelan}, A.~M., {Sip{\H{o}}cz}, B.~M., {et~al.} 2018, \aj, 156, 123

\bibitem[{{Bennett} \& {Bovy}(2019)}]{Bennett_Bovy_19_MNRAS}
{Bennett}, M. \& {Bovy}, J. 2019, \mnras, 482, 1417

\bibitem[{{Bialy} {et~al.}(2021){Bialy}, {Zucker}, {Goodman}, {Foley}, {Alves}, {Semenov}, {Benjamin}, {Leike}, \& {En{\ss}lin}}]{Bialy_21_ApJ}
{Bialy}, S., {Zucker}, C., {Goodman}, A., {et~al.} 2021, \apjl, 919, L5

\bibitem[{{Bisbas} {et~al.}(2011){Bisbas}, {W{\"u}nsch}, {Whitworth}, {Hubber}, \& {Walch}}]{Bisbas_11_ApJ}
{Bisbas}, T.~G., {W{\"u}nsch}, R., {Whitworth}, A.~P., {Hubber}, D.~A., \& {Walch}, S. 2011, \apj, 736, 142

\bibitem[{{Blaauw}(1946)}]{Blaauw_46_PGro}
{Blaauw}, A. 1946, Publications of the Kapteyn Astronomical Laboratory Groningen, 52, 1

\bibitem[{{Blaauw}(1964)}]{Blaauw_64_ARAA}
{Blaauw}, A. 1964, \araa, 2, 213

\bibitem[{Bobylev \& Bajkova(2024)}]{Bobylev_24_arxiv}
Bobylev, V.~V. \& Bajkova, A.~T. 2024, Research in Astronomy and Astrophysics, 24, 055004

\bibitem[{{Boulanger} {et~al.}(1998){Boulanger}, {Bronfman}, {Dame}, \& {Thaddeus}}]{Boulanger_98_AA}
{Boulanger}, F., {Bronfman}, L., {Dame}, T.~M., \& {Thaddeus}, P. 1998, \aap, 332, 273

\bibitem[{Bouy \& Alves(2015)}]{Bouy2015-ce}
Bouy, H. \& Alves, J. 2015, Astron. Astrophys., 584, A26

\bibitem[{Bovy(2015)}]{galpy_15}
Bovy, J. 2015, The Astrophysical Journal Supplement Series, 216, 29

\bibitem[{{Bressan} {et~al.}(2012){Bressan}, {Marigo}, {Girardi}, {Salasnich}, {Dal Cero}, {Rubele}, \& {Nanni}}]{Bressan_12_MNRAS}
{Bressan}, A., {Marigo}, P., {Girardi}, L., {et~al.} 2012, \mnras, 427, 127

\bibitem[{{Brice{\~n}o-Morales} \& {Chanam{\'e}}(2023)}]{Briceno_23_MNRAS}
{Brice{\~n}o-Morales}, G. \& {Chanam{\'e}}, J. 2023, \mnras, 522, 1288

\bibitem[{{Buder} {et~al.}(2021){Buder}, {Sharma}, {Kos}, {Amarsi}, {Nordlander}, {Lind}, {Martell}, {Asplund}, {Bland-Hawthorn}, {Casey}, {de Silva}, {D'Orazi}, {Freeman}, {Hayden}, {Lewis}, {Lin}, {Schlesinger}, {Simpson}, {Stello}, {Zucker}, {Zwitter}, {Beeson}, {Buck}, {Casagrande}, {Clark}, {{\v{C}}otar}, {da Costa}, {de Grijs}, {Feuillet}, {Horner}, {Kafle}, {Khanna}, {Kobayashi}, {Liu}, {Montet}, {Nandakumar}, {Nataf}, {Ness}, {Spina}, {Tepper-Garc{\'\i}a}, {Ting}, {Traven}, {Vogrin{\v{c}}i{\v{c}}}, {Wittenmyer}, {Wyse}, {{\v{Z}}erjal}, \& {Galah Collaboration}}]{Galah_21_MNRAS}
{Buder}, S., {Sharma}, S., {Kos}, J., {et~al.} 2021, \mnras, 506, 150

\bibitem[{{Casanova} {et~al.}(1995){Casanova}, {Montmerle}, {Feigelson}, \& {Andre}}]{Casanova_95_ApJ}
{Casanova}, S., {Montmerle}, T., {Feigelson}, E.~D., \& {Andre}, P. 1995, \apj, 439, 752

\bibitem[{{Chen} {et~al.}(2015){Chen}, {Bressan}, {Girardi}, {Marigo}, {Kong}, \& {Lanza}}]{Chen_15_MNRAS}
{Chen}, Y., {Bressan}, A., {Girardi}, L., {et~al.} 2015, \mnras, 452, 1068

\bibitem[{{Covino} {et~al.}(1997){Covino}, {Alcala}, {Allain}, {Bouvier}, {Terranegra}, \& {Krautter}}]{Covino_97_AA}
{Covino}, E., {Alcala}, J.~M., {Allain}, S., {et~al.} 1997, \aap, 328, 187

\bibitem[{{Dale} {et~al.}(2007){Dale}, {Clark}, \& {Bonnell}}]{Dale_07_MNRAS}
{Dale}, J.~E., {Clark}, P.~C., \& {Bonnell}, I.~A. 2007, \mnras, 377, 535

\bibitem[{{Dale} {et~al.}(2012){Dale}, {Ercolano}, \& {Bonnell}}]{Dale_12_MNRAS}
{Dale}, J.~E., {Ercolano}, B., \& {Bonnell}, I.~A. 2012, \mnras, 427, 2852

\bibitem[{{Dame} {et~al.}(2001){Dame}, {Hartmann}, \& {Thaddeus}}]{Dame_01_ApJ}
{Dame}, T.~M., {Hartmann}, D., \& {Thaddeus}, P. 2001, \apj, 547, 792

\bibitem[{{Damiani} {et~al.}(2019){Damiani}, {Prisinzano}, {Pillitteri}, {Micela}, \& {Sciortino}}]{Damiani_19_AA}
{Damiani}, F., {Prisinzano}, L., {Pillitteri}, I., {Micela}, G., \& {Sciortino}, S. 2019, \aap, 623, A112

\bibitem[{{de Geus}(1992)}]{deGeus_92_AA}
{de Geus}, E.~J. 1992, \aap, 262, 258

\bibitem[{{de Geus} {et~al.}(1989){de Geus}, {de Zeeuw}, \& {Lub}}]{deGeus_89_AA}
{de Geus}, E.~J., {de Zeeuw}, P.~T., \& {Lub}, J. 1989, \aap, 216, 44

\bibitem[{{de Zeeuw} {et~al.}(1999){de Zeeuw}, {Hoogerwerf}, {de Bruijne}, {Brown}, \& {Blaauw}}]{deZeeuw_99_AJ}
{de Zeeuw}, P.~T., {Hoogerwerf}, R., {de Bruijne}, J.~H.~J., {Brown}, A.~G.~A., \& {Blaauw}, A. 1999, \aj, 117, 354

\bibitem[{{Deb} {et~al.}(2018){Deb}, {Kothes}, \& {Rosolowsky}}]{Deb_18_MNRAS}
{Deb}, S., {Kothes}, R., \& {Rosolowsky}, E. 2018, \mnras, 481, 1862

\bibitem[{{Dickson-Vandervelde} {et~al.}(2021){Dickson-Vandervelde}, {Wilson}, \& {Kastner}}]{Dickson_21_AJ}
{Dickson-Vandervelde}, D.~A., {Wilson}, E.~C., \& {Kastner}, J.~H. 2021, \aj, 161, 87

\bibitem[{{Edenhofer} {et~al.}(2024b){Edenhofer}, {Alves}, {Zucker}, {Posch}, \& {En{\ss}lin}}]{Edenhofer_24b_AA}
{Edenhofer}, G., {Alves}, J., {Zucker}, C., {Posch}, L., \& {En{\ss}lin}, T.~A. 2024b, \aap, 687, L9

\bibitem[{{Edenhofer} {et~al.}(2024a){Edenhofer}, {Zucker}, {Frank}, {Saydjari}, {Speagle}, {Finkbeiner}, \& {En{\ss}lin}}]{Edenhofer_24a_AA}
{Edenhofer}, G., {Zucker}, C., {Frank}, P., {et~al.} 2024a, \aap, 685, A82

\bibitem[{{Elmegreen}(2011)}]{Elmegreen_11_EAS}
{Elmegreen}, B.~G. 2011, in EAS Publications Series, Vol.~51, EAS Publications Series, ed. C.~{Charbonnel} \& T.~{Montmerle}, 45--58

\bibitem[{{Elmegreen} \& {Lada}(1977)}]{Elmegreen_Lada_77_ApJ}
{Elmegreen}, B.~G. \& {Lada}, C.~J. 1977, \apj, 214, 725

\bibitem[{{Fern{\'a}ndez} {et~al.}(2008){Fern{\'a}ndez}, {Figueras}, \& {Torra}}]{Fernandez_08_AA}
{Fern{\'a}ndez}, D., {Figueras}, F., \& {Torra}, J. 2008, \aap, 480, 735

\bibitem[{{Foley} {et~al.}(2023){Foley}, {Goodman}, {Zucker}, {Forbes}, {Konietzka}, {Swiggum}, {Alves}, {Bally}, {Soler}, {Gro{\ss}schedl}, {Bialy}, {Grudi{\'c}}, {Leike}, \& {En{\ss}lin}}]{Foley_23_ApJ}
{Foley}, M.~M., {Goodman}, A., {Zucker}, C., {et~al.} 2023, \apj, 947, 66

\bibitem[{{Franchini} {et~al.}(1992){Franchini}, {Covino}, {Stalio}, {Terranegra}, \& {Chavarria-K.}}]{Franchini_92_AA}
{Franchini}, M., {Covino}, E., {Stalio}, R., {Terranegra}, L., \& {Chavarria-K.}, C. 1992, \aap, 256, 525

\bibitem[{{Fuchs} {et~al.}(2006){Fuchs}, {Breitschwerdt}, {de Avillez}, {Dettbarn}, \& {Flynn}}]{Fuchs_06_MNRAS}
{Fuchs}, B., {Breitschwerdt}, D., {de Avillez}, M.~A., {Dettbarn}, C., \& {Flynn}, C. 2006, \mnras, 373, 993

\bibitem[{{Gaczkowski} {et~al.}(2017){Gaczkowski}, {Roccatagliata}, {Flaischlen}, {Kr{\"o}ll}, {Krause}, {Burkert}, {Diehl}, {Fierlinger}, {Ngoumou}, \& {Preibisch}}]{Gaczkowski_17_AA}
{Gaczkowski}, B., {Roccatagliata}, V., {Flaischlen}, S., {et~al.} 2017, \aap, 608, A102

\bibitem[{{Gaia Collaboration} {et~al.}(2018){Gaia Collaboration}, {Brown}, {Vallenari}, {Prusti}, {de Bruijne}, {Babusiaux}, \& {Bailer-Jones}}]{Gaia_18_AA}
{Gaia Collaboration}, {Brown}, A.~G.~A., {Vallenari}, A., {et~al.} 2018, \aap, 616, A1

\bibitem[{{Gaia Collaboration} {et~al.}(2016){Gaia Collaboration}, {Prusti}, {de Bruijne}, {Brown}, {Vallenari}, {Babusiaux}, \& {Bailer-Jones}}]{Gaia_16_AA}
{Gaia Collaboration}, {Prusti}, T., {de Bruijne}, J.~H.~J., {et~al.} 2016, \aap, 595, A1

\bibitem[{{Gaia Collaboration} {et~al.}(2023){Gaia Collaboration}, {Valinieri}, {Brown}, {Prusti}, {de Bruijne}, {Arenou}, {Babusiaux}, \& {Biermann}}]{Gaia_23_AA}
{Gaia Collaboration}, {Valinieri}, A., {Brown}, A.~G.~A., {et~al.} 2023, \aap, 674, A1

\bibitem[{{Getman} {et~al.}(2007){Getman}, {Feigelson}, {Garmire}, {Broos}, \& {Wang}}]{Getman_07_ApJ}
{Getman}, K.~V., {Feigelson}, E.~D., {Garmire}, G., {Broos}, P., \& {Wang}, J. 2007, \apj, 654, 316

\bibitem[{{Getman} {et~al.}(2009){Getman}, {Feigelson}, {Luhman}, {Sicilia-Aguilar}, {Wang}, \& {Garmire}}]{Getman_09_ApJ}
{Getman}, K.~V., {Feigelson}, E.~D., {Luhman}, K.~L., {et~al.} 2009, \apj, 699, 1454

\bibitem[{{Getman} {et~al.}(2012){Getman}, {Feigelson}, {Sicilia-Aguilar}, {Broos}, {Kuhn}, \& {Garmire}}]{Getman_12_MNRAS}
{Getman}, K.~V., {Feigelson}, E.~D., {Sicilia-Aguilar}, A., {et~al.} 2012, \mnras, 426, 2917

\bibitem[{{Girichidis} {et~al.}(2016){Girichidis}, {Walch}, {Naab}, {Gatto}, {W{\"u}nsch}, {Glover}, {Klessen}, {Clark}, {Peters}, {Derigs}, \& {Baczynski}}]{Girichidis_16_MNRAS}
{Girichidis}, P., {Walch}, S., {Naab}, T., {et~al.} 2016, \mnras, 456, 3432

\bibitem[{{Goldman} {et~al.}(2018){Goldman}, {R{\"o}ser}, {Schilbach}, {Mo{\'o}r}, \& {Henning}}]{Goldman_18_ApJ}
{Goldman}, B., {R{\"o}ser}, S., {Schilbach}, E., {Mo{\'o}r}, A.~C., \& {Henning}, T. 2018, \apj, 868, 32

\bibitem[{{Grasser} {et~al.}(2021){Grasser}, {Ratzenb{\"o}ck}, {Alves}, {Gro{\ss}schedl}, {Meingast}, {Zucker}, {Hacar}, {Lada}, {Goodman}, {Lombardi}, {Forbes}, {Bomze}, \& {M{\"o}ller}}]{Grasser_21_AA}
{Grasser}, N., {Ratzenb{\"o}ck}, S., {Alves}, J., {et~al.} 2021, \aap, 652, A2

\bibitem[{{GRAVITY Collaboration} {et~al.}(2018){GRAVITY Collaboration}, {Abuter}, {Amorim}, {Anugu}, {Baub{\"o}ck}, {Benisty}, {Berger}, {Blind}, \& {Bonnet}}]{GravityColl_18_AA}
{GRAVITY Collaboration}, {Abuter}, R., {Amorim}, A., {et~al.} 2018, \aap, 615, L15

\bibitem[{{Gro{\ss}schedl} {et~al.}(2021){Gro{\ss}schedl}, {Alves}, {Meingast}, \& {Herbst-Kiss}}]{Grossschedl_21_AA}
{Gro{\ss}schedl}, J.~E., {Alves}, J., {Meingast}, S., \& {Herbst-Kiss}, G. 2021, \aap, 647, A91

\bibitem[{{Hacar} {et~al.}(2016){Hacar}, {Alves}, {Forbrich}, {Meingast}, {Kubiak}, \& {Gro{\ss}schedl}}]{Hacar_16_AA}
{Hacar}, A., {Alves}, J., {Forbrich}, J., {et~al.} 2016, \aap, 589, A80

\bibitem[{Harris {et~al.}(2020)Harris, Millman, van~der Walt, Gommers, Virtanen, Cournapeau, Wieser, Taylor, Berg, Smith, Kern, Picus, Hoyer, van Kerkwijk, Brett, Haldane, del R{\'{i}}o, Wiebe, Peterson, G{\'{e}}rard-Marchant, Sheppard, Reddy, Weckesser, Abbasi, Gohlke, \& Oliphant}]{numpy_20}
Harris, C.~R., Millman, K.~J., van~der Walt, S.~J., {et~al.} 2020, Nature, 585, 357

\bibitem[{{Hartigan} {et~al.}(2015){Hartigan}, {Reiter}, {Smith}, \& {Bally}}]{Hartigan_15_AJ}
{Hartigan}, P., {Reiter}, M., {Smith}, N., \& {Bally}, J. 2015, \aj, 149, 101

\bibitem[{{Hendler} {et~al.}(2018){Hendler}, {Pinilla}, {Pascucci}, {Pohl}, {Mulders}, {Henning}, {Dong}, {Clarke}, {Owen}, \& {Hollenbach}}]{Hendler_18_MNRAS}
{Hendler}, N.~P., {Pinilla}, P., {Pascucci}, I., {et~al.} 2018, \mnras, 475, L62

\bibitem[{{Herrington} {et~al.}(2023){Herrington}, {Dobbs}, \& {Bending}}]{Herrington_23_MNRAS}
{Herrington}, N.~P., {Dobbs}, C.~L., \& {Bending}, T. J.~R. 2023, \mnras, 521, 5712

\bibitem[{{HI4PI Collaboration}(2016)}]{Hi4Pi_16_AA}
{HI4PI Collaboration}. 2016, \aap, 594, A116

\bibitem[{{Hosokawa} \& {Inutsuka}(2006)}]{Hosokawa_Inutsuka_06_ApJ}
{Hosokawa}, T. \& {Inutsuka}, S.-i. 2006, \apj, 646, 240

\bibitem[{Hunter(2007)}]{matplotlib_07}
Hunter, J.~D. 2007, Computing in Science \& Engineering, 9, 90

\bibitem[{Inc.(2015)}]{plotly}
Inc., P.~T. 2015, Collaborative data science

\bibitem[{{Inutsuka} {et~al.}(2015){Inutsuka}, {Inoue}, {Iwasaki}, \& {Hosokawa}}]{Inutsuka_15_AA}
{Inutsuka}, S.-i., {Inoue}, T., {Iwasaki}, K., \& {Hosokawa}, T. 2015, \aap, 580, A49

\bibitem[{{Katz} {et~al.}(2023){Katz}, {Sartoretti}, {Guerrier}, {Panuzzo}, {Seabroke}, {Th{\'e}venin}, {Cropper}, {Benson}, {Blomme}, {Haigron}, {Marchal}, {Smith}, {Baker}, {Chemin}, {Damerdji}, {David}, {Dolding}, {Fr{\'e}mat}, {Gosset}, {Jan{\ss}en}, {Jasniewicz}, {Lobel}, {Plum}, {Samaras}, {Snaith}, {Soubiran}, {Vanel}, {Zwitter}, {Antoja}, {Arenou}, {Babusiaux}, {Brouillet}, {Caffau}, {Di Matteo}, {Fabre}, {Fabricius}, {Fragkoudi}, {Haywood}, {Huckle}, {Hottier}, {Lasne}, {Leclerc}, {Mastrobuono-Battisti}, {Royer}, {Teyssier}, {Zorec}, {Crifo}, {Jean-Antoine Piccolo}, {Turon}, \& {Viala}}]{Katz_23_AA}
{Katz}, D., {Sartoretti}, P., {Guerrier}, A., {et~al.} 2023, \aap, 674, A5

\bibitem[{{Kendrew} {et~al.}(2012){Kendrew}, {Simpson}, {Bressert}, {Povich}, {Sherman}, {Lintott}, {Robitaille}, {Schawinski}, \& {Wolf-Chase}}]{Kendrew_12_ApJ}
{Kendrew}, S., {Simpson}, R., {Bressert}, E., {et~al.} 2012, \apj, 755, 71

\bibitem[{{Kerr} \& {Lynden-Bell}(1986)}]{Kerr_LyndenBell_86_HiA}
{Kerr}, F.~J. \& {Lynden-Bell}, D. 1986, Highlights of Astronomy, 7, 889

\bibitem[{{Kerr} {et~al.}(2024){Kerr}, {Kraus}, {Krolikowski}, {Bouma}, \& {Farias}}]{Kerr_24_arXiv}
{Kerr}, R., {Kraus}, A.~L., {Krolikowski}, D., {Bouma}, L.~G., \& {Farias}, J.~P. 2024, \apj, accepted, arXiv:2406.19530

\bibitem[{{Kerr} {et~al.}(2021){Kerr}, {Rizzuto}, {Kraus}, \& {Offner}}]{Kerr_21_ApJ}
{Kerr}, R. M.~P., {Rizzuto}, A.~C., {Kraus}, A.~L., \& {Offner}, S. S.~R. 2021, \apj, 917, 23

\bibitem[{{Klein} {et~al.}(1980){Klein}, {Sandford}, \& {Whitaker}}]{Klein_80_SSRv}
{Klein}, R.~I., {Sandford}, M.~T., I., \& {Whitaker}, R.~W. 1980, \ssr, 27, 275

\bibitem[{Konietzka {et~al.}(2024)Konietzka, Goodman, Zucker, Burkert, Alves, Foley, Swiggum, Koller, \& Miret-Roig}]{Konietzka_24_Nature}
Konietzka, R., Goodman, A.~A., Zucker, C., {et~al.} 2024, \nat, 628, 62

\bibitem[{{Kos} {et~al.}(2021){Kos}, {Bland-Hawthorn}, {Buder}, {Nordlander}, {Spina}, {Beeson}, {Lind}, {Asplund}, {Freeman}, {Hayden}, {Lewis}, {Martell}, {Sharma}, {De Silva}, {Simpson}, {Zucker}, {Zwitter}, {{\v{C}}otar}, {Horner}, {Ting}, \& {Traven}}]{Kos_21_MNRAS}
{Kos}, J., {Bland-Hawthorn}, J., {Buder}, S., {et~al.} 2021, \mnras, 506, 4232

\bibitem[{{Krause} {et~al.}(2018){Krause}, {Burkert}, {Diehl}, {Fierlinger}, {Gaczkowski}, {Kroell}, {Ngoumou}, {Roccatagliata}, {Siegert}, \& {Preibisch}}]{Krause_18_AA}
{Krause}, M. G.~H., {Burkert}, A., {Diehl}, R., {et~al.} 2018, \aap, 619, A120

\bibitem[{{Kroupa}(2001)}]{Kroupa_01_MNRAS}
{Kroupa}, P. 2001, \mnras, 322, 231

\bibitem[{{Lada} {et~al.}(2012){Lada}, {Forbrich}, {Lombardi}, \& {Alves}}]{Lada_12_ApJ}
{Lada}, C.~J., {Forbrich}, J., {Lombardi}, M., \& {Alves}, J.~F. 2012, \apj, 745, 190

\bibitem[{{Ladjelate} {et~al.}(2020){Ladjelate}, {Andr{\'e}}, {K{\"o}nyves}, {Ward-Thompson}, {Men'shchikov}, {Bracco}, {Palmeirim}, {Roy}, {Shimajiri}, {Kirk}, {Arzoumanian}, {Benedettini}, {Di Francesco}, {Fiorellino}, {Schneider}, {Pezzuto}, {Motte}, \& {Herschel Gould Belt Survey Team}}]{Ladjelate_20_AA}
{Ladjelate}, B., {Andr{\'e}}, P., {K{\"o}nyves}, V., {et~al.} 2020, \aap, 638, A74

\bibitem[{{Lawson} {et~al.}(2009){Lawson}, {Lyo}, \& {Bessell}}]{Lawson_09_MNRAS}
{Lawson}, W.~A., {Lyo}, A.~R., \& {Bessell}, M.~S. 2009, \mnras, 400, L29

\bibitem[{{Leike} {et~al.}(2020){Leike}, {Glatzle}, \& {En{\ss}lin}}]{Leike_20_AA}
{Leike}, R.~H., {Glatzle}, M., \& {En{\ss}lin}, T.~A. 2020, \aap, 639, A138

\bibitem[{{Lombardi} {et~al.}(2008){Lombardi}, {Lada}, \& {Alves}}]{Lombardi_08b_AA}
{Lombardi}, M., {Lada}, C.~J., \& {Alves}, J. 2008, \aap, 489, 143

\bibitem[{{Lopez Mart{\'\i}} {et~al.}(2013){Lopez Mart{\'\i}}, {Jimenez Esteban}, {Bayo}, {Barrado}, {Solano}, \& {Rodrigo}}]{LopezMartin_13_AA}
{Lopez Mart{\'\i}}, B., {Jimenez Esteban}, F., {Bayo}, A., {et~al.} 2013, \aap, 551, A46

\bibitem[{{Malinen} {et~al.}(2014){Malinen}, {Juvela}, {Zahorecz}, {Rivera-Ingraham}, {Montillaud}, {Arimatsu}, {Bernard}, {Doi}, {Haikala}, {Kawabe}, {Marton}, {McGehee}, {Pelkonen}, {Ristorcelli}, {Shimajiri}, {Takita}, {T{\'o}th}, {Tsukagoshi}, \& {Ysard}}]{Malinen_14_AA}
{Malinen}, J., {Juvela}, M., {Zahorecz}, S., {et~al.} 2014, \aap, 563, A125

\bibitem[{{Marigo} {et~al.}(2017){Marigo}, {Girardi}, {Bressan}, {Rosenfield}, {Aringer}, {Chen}, {Dussin}, {Nanni}, {Pastorelli}, {Rodrigues}, {Trabucchi}, {Bladh}, {Dalcanton}, {Groenewegen}, {Montalb{\'a}n}, \& {Wood}}]{Marigo_17_ApJ}
{Marigo}, P., {Girardi}, L., {Bressan}, A., {et~al.} 2017, \apj, 835, 77

\bibitem[{{Marshall} \& {Kerton}(2019)}]{Marshall_19_MNRAS}
{Marshall}, B. \& {Kerton}, C.~R. 2019, \mnras, 489, 4809

\bibitem[{McCray(1983)}]{McCray_83}
McCray, R. 1983, Highlights of Astronomy, 6, 565

\bibitem[{{Megeath} {et~al.}(1996){Megeath}, {Cox}, {Bronfman}, \& {Roelfsema}}]{Megeath_96_AA}
{Megeath}, S.~T., {Cox}, P., {Bronfman}, L., \& {Roelfsema}, P.~R. 1996, \aap, 305, 296

\bibitem[{{Megeath} \& {Wilson}(1997)}]{Megeath_97_AJ}
{Megeath}, S.~T. \& {Wilson}, T.~L. 1997, \aj, 114, 1106

\bibitem[{{Miao} {et~al.}(2006){Miao}, {White}, {Nelson}, {Thompson}, \& {Morgan}}]{Miao_06_MNRAS}
{Miao}, J., {White}, G.~J., {Nelson}, R., {Thompson}, M., \& {Morgan}, L. 2006, \mnras, 369, 143

\bibitem[{{Miret-Roig} {et~al.}(2024){Miret-Roig}, {Alves}, {Ratzenb{\"o}ck}, {Galli}, {Bouy}, \& {Figueras}}]{MiretRoig_24_AA}
{Miret-Roig}, N., {Alves}, J., {Ratzenb{\"o}ck}, S., {et~al.} 2024, \aap, submitted

\bibitem[{{Miret-Roig} {et~al.}(2022){Miret-Roig}, {Galli}, {Olivares}, {Bouy}, {Alves}, \& {Barrado}}]{MiretRoig_22_AA}
{Miret-Roig}, N., {Galli}, P.~A.~B., {Olivares}, J., {et~al.} 2022, \aap, 667, A163

\bibitem[{{Mizuno} {et~al.}(2001){Mizuno}, {Yamaguchi}, {Tachihara}, {Toyoda}, {Aoyama}, {Yamamoto}, {Onishi}, \& {Fukui}}]{Mizuno_01_PASJ}
{Mizuno}, A., {Yamaguchi}, R., {Tachihara}, K., {et~al.} 2001, Publications of the Astronomical Society of Japan, 53, 1071

\bibitem[{{Murphy} {et~al.}(2013){Murphy}, {Lawson}, \& {Bessell}}]{Murphy_13_MNRAS}
{Murphy}, S.~J., {Lawson}, W.~A., \& {Bessell}, M.~S. 2013, \mnras, 435, 1325

\bibitem[{{Nehm{\'e}} {et~al.}(2008){Nehm{\'e}}, {Gry}, {Boulanger}, {Le Bourlot}, {Pineau Des For{\^e}ts}, \& {Falgarone}}]{Nehme_08_AA}
{Nehm{\'e}}, C., {Gry}, C., {Boulanger}, F., {et~al.} 2008, \aap, 483, 471

\bibitem[{{Neuh{\"a}user} \& {Forbrich}(2008)}]{Neuhauser_Forbich_08}
{Neuh{\"a}user}, R. \& {Forbrich}, J. 2008, in Handbook of Star Forming Regions, Volume II, ed. B.~{Reipurth}, Vol.~5, 735

\bibitem[{{Neuh{\"a}user} {et~al.}(2020){Neuh{\"a}user}, {Gie{\ss}ler}, \& {Hambaryan}}]{Neuhauser_20_MNRAS}
{Neuh{\"a}user}, R., {Gie{\ss}ler}, F., \& {Hambaryan}, V.~V. 2020, \mnras, 498, 899

\bibitem[{{O'Neill} {et~al.}(2024){O'Neill}, {Zucker}, {Goodman}, \& {Edenhofer}}]{ONeill_24_ApJ}
{O'Neill}, T.~J., {Zucker}, C., {Goodman}, A.~A., \& {Edenhofer}, G. 2024, \apj, 973, 136

\bibitem[{{Pang} {et~al.}(2021){Pang}, {Yu}, {Tang}, {Hong}, {Yuan}, {Pasquato}, \& {Kouwenhoven}}]{Pang_21_ApJ}
{Pang}, X., {Yu}, Z., {Tang}, S.-Y., {et~al.} 2021, \apj, 923, 20

\bibitem[{Pecaut \& Mamajek(2016)}]{Pecault_Mamajek_16_MNRAS}
Pecaut, M.~J. \& Mamajek, E.~E. 2016, Monthly Notices of the Royal Astronomical Society, 461, 794

\bibitem[{{Piecka} {et~al.}(2024){Piecka}, {Hutschenreuter}, \& {Alves}}]{Piecka_24_AA}
{Piecka}, M., {Hutschenreuter}, S., \& {Alves}, J. 2024, \aap, 689, A84

\bibitem[{{Planck Collaboration XI}(2014)}]{Planck_14_AA}
{Planck Collaboration XI}. 2014, A\&A, 571, A11

\bibitem[{{Planck Collaboration XI}(2020)}]{Planck_20_AA}
{Planck Collaboration XI}. 2020, \aap, 641, A1

\bibitem[{{Posch} {et~al.}(2023){Posch}, {Miret-Roig}, {Alves}, {Ratzenb{\"o}ck}, {Gro{\ss}schedl}, {Meingast}, {Zucker}, \& {Burkert}}]{Posch_23_AA}
{Posch}, L., {Miret-Roig}, N., {Alves}, J., {et~al.} 2023, \aap, 679, L10

\bibitem[{{Preibisch} \& {Mamajek}(2008)}]{Preibisch_08_book}
{Preibisch}, T. \& {Mamajek}, E. 2008, in Handbook of Star Forming Regions, Volume II, ed. B.~{Reipurth}, Vol.~5, 235

\bibitem[{{Preibisch} \& {Zinnecker}(1999)}]{Preibisch_99_AJ}
{Preibisch}, T. \& {Zinnecker}, H. 1999, \aj, 117, 2381

\bibitem[{{Ratzenb{\"o}ck} {et~al.}(2023b){Ratzenb{\"o}ck}, {Gro{\ss}schedl}, {Alves}, {Miret-Roig}, {Bomze}, {Forbes}, {Goodman}, {Hacar}, {Lin}, {Meingast}, {M{\"o}ller}, {Piecka}, {Posch}, {Rottensteiner}, {Swiggum}, \& {Zucker}}]{Ratzenboeck_23b_AA}
{Ratzenb{\"o}ck}, S., {Gro{\ss}schedl}, J.~E., {Alves}, J., {et~al.} 2023b, \aap, 678, A71

\bibitem[{{Ratzenb{\"o}ck} {et~al.}(2023a){Ratzenb{\"o}ck}, {Gro{\ss}schedl}, {M{\"o}ller}, {Alves}, {Bomze}, \& {Meingast}}]{Ratzenboeck_23a_AA}
{Ratzenb{\"o}ck}, S., {Gro{\ss}schedl}, J.~E., {M{\"o}ller}, T., {et~al.} 2023a, \aap, 677, A59

\bibitem[{{Redfield} \& {Linsky}(2008)}]{Redfield_Linsky_08_ApJ}
{Redfield}, S. \& {Linsky}, J.~L. 2008, \apj, 673, 283

\bibitem[{{Roccatagliata} {et~al.}(2013){Roccatagliata}, {Preibisch}, {Ratzka}, \& {Gaczkowski}}]{Roccatagliata_13_AA}
{Roccatagliata}, V., {Preibisch}, T., {Ratzka}, T., \& {Gaczkowski}, B. 2013, \aap, 554, A6

\bibitem[{{Rogers} \& {Pittard}(2013)}]{Rogers_Pittard_13_MNRAS}
{Rogers}, H. \& {Pittard}, J.~M. 2013, \mnras, 431, 1337

\bibitem[{{Rogers} \& {Pittard}(2014)}]{Rogers_Pittard_14_MNRAS}
{Rogers}, H. \& {Pittard}, J.~M. 2014, \mnras, 441, 964

\bibitem[{{Sabatini} {et~al.}(2024){Sabatini}, {Podio}, {Codella}, {Watanabe}, {De Simone}, {Bianchi}, {Ceccarelli}, {Chandler}, {Sakai}, {Svoboda}, {Testi}, {Aikawa}, {Balucani}, {Bouvier}, {Caselli}, {Caux}, {Chahine}, {Charnley}, {Cuello}, {Dulieu}, {Evans}, {Fedele}, {Feng}, {Fontani}, {Hama}, {Hanawa}, {Herbst}, {Hirota}, {Isella}, {J{\'\i}menez-Serra}, {Johnstone}, {Lefloch}, {Le Gal}, {Loinard}, {Liu}, {L{\'o}pez-Sepulcre}, {Maud}, {Maureira}, {Menard}, {Miotello}, {Moellenbrock}, {Nomura}, {Oba}, {Ohashi}, {Okoda}, {Oya}, {Pineda}, {Rimola}, {Sakai}, {Segura-Cox}, {Shirley}, {Vastel}, {Viti}, {Watanabe}, {Zhang}, {Zhang}, \& {Yamamoto}}]{Sabatini_24_AA}
{Sabatini}, G., {Podio}, L., {Codella}, C., {et~al.} 2024, \aap, 684, L12

\bibitem[{Salvatier {et~al.}(2016)Salvatier, Wiecki, \& Fonnesbeck}]{PyMC_16}
Salvatier, J., Wiecki, T.~V., \& Fonnesbeck, C. 2016, {PeerJ} Computer Science, 2, e55

\bibitem[{{Sandell} {et~al.}(2021){Sandell}, {Reipurth}, {Vacca}, \& {Bajaj}}]{Sandell_21_ApJ}
{Sandell}, G., {Reipurth}, B., {Vacca}, W.~D., \& {Bajaj}, N.~S. 2021, \apj, 920, 7

\bibitem[{{Schisano} {et~al.}(2009){Schisano}, {Covino}, {Alcal{\'a}}, {Esposito}, {Gandolfi}, \& {Guenther}}]{Schisano_09_AA}
{Schisano}, E., {Covino}, E., {Alcal{\'a}}, J.~M., {et~al.} 2009, \aap, 501, 1013

\bibitem[{{Schmidt}(1959)}]{Schmidt_59_ApJ}
{Schmidt}, M. 1959, \apj, 129, 243

\bibitem[{{Sch{\"o}nrich} {et~al.}(2010){Sch{\"o}nrich}, {Binney}, \& {Dehnen}}]{Schoenrich_10_MNRAS}
{Sch{\"o}nrich}, R., {Binney}, J., \& {Dehnen}, W. 2010, \mnras, 403, 1829

\bibitem[{{Snider} {et~al.}(2009){Snider}, {Hester}, {Desch}, {Healy}, \& {Bally}}]{Snider_09_ApJ}
{Snider}, K.~D., {Hester}, J.~J., {Desch}, S.~J., {Healy}, K.~R., \& {Bally}, J. 2009, \apj, 700, 506

\bibitem[{Spina {et~al.}(2017)Spina, Randich, Magrini, Jeffries, Friel, Sacco, Pancino, Bonito, Bravi, Franciosini, Klutsch, Montes, Gilmore, Vallenari, Bensby, Bragaglia, Flaccomio, Koposov, Korn, Lanzafame, Smiljanic, Bayo, Carraro, Casey, Costado, Damiani, Donati, Frasca, Hourihane, Jofré, Lewis, Lind, Monaco, Morbidelli, Prisinzano, Sousa, Worley, \& Zaggia}]{Spina2017-xl}
Spina, L., Randich, S., Magrini, L., {et~al.} 2017, Astron. Astrophys. Suppl. Ser., 601, A70

\bibitem[{{Squicciarini} {et~al.}(2021){Squicciarini}, {Gratton}, {Bonavita}, \& {Mesa}}]{Squicciarini_21_MNRAS}
{Squicciarini}, V., {Gratton}, R., {Bonavita}, M., \& {Mesa}, D. 2021, \mnras, 507, 1381

\bibitem[{Swiggum {et~al.}(2024)Swiggum, Alves, Benjamin, Ratzenböck, Miret-Roig, Großschedl, Meingast, Goodman, Konietzka, Zucker, Hunt, \& Reffert}]{Swiggum_24_Nature}
Swiggum, C., Alves, J., Benjamin, R., {et~al.} 2024, Nature

\bibitem[{{Tachihara} {et~al.}(2024){Tachihara}, {Fukaya}, {Tokuda}, {Yamasaki}, {Nishioka}, {Abe}, {Inoue}, {Harada}, {Shoshi}, {Nozaki}, {Sato}, {Omura}, {Fujishiro}, {Fukagawa}, {Machida}, {Kanai}, {Oasa}, {Onishi}, {Saigo}, \& {Fukui}}]{Tachihara_24_ApJ}
{Tachihara}, K., {Fukaya}, N., {Tokuda}, K., {et~al.} 2024, \apj, 968, 131

\bibitem[{{Taylor} \& {Storey}(1984)}]{Taylor_Storey_84_MNRAS}
{Taylor}, K.~N.~R. \& {Storey}, J.~W.~V. 1984, \mnras, 209, 5P

\bibitem[{{Thompson} {et~al.}(2012){Thompson}, {Urquhart}, {Moore}, \& {Morgan}}]{Thompson_12_MNRAS}
{Thompson}, M.~A., {Urquhart}, J.~S., {Moore}, T.~J.~T., \& {Morgan}, L.~K. 2012, \mnras, 421, 408

\bibitem[{{Tobin} {et~al.}(2009){Tobin}, {Hartmann}, {Furesz}, {Mateo}, \& {Megeath}}]{Tobin_09_ApJ}
{Tobin}, J.~J., {Hartmann}, L., {Furesz}, G., {Mateo}, M., \& {Megeath}, S.~T. 2009, \apj, 697, 1103

\bibitem[{{Varga} {et~al.}(2024){Varga}, {Kastner}, {Dickson-Vandervelde}, \& {Binks}}]{Varga_24_arXiv}
{Varga}, A., {Kastner}, J.~H., {Dickson-Vandervelde}, D.~A., \& {Binks}, A.~S. 2024, \aj, accepted, arXiv:2409.17521

\bibitem[{Virtanen {et~al.}(2020)Virtanen, Gommers, Oliphant, Haberland, Reddy, Cournapeau, Burovski, Peterson, Weckesser, Bright, {van der Walt}, Brett, Wilson, Millman, Mayorov, Nelson, Jones, Kern, Larson, Carey, Polat, Feng, Moore, {VanderPlas}, Laxalde, Perktold, Cimrman, Henriksen, Quintero, Harris, Archibald, Ribeiro, Pedregosa, {van Mulbregt}, \& {SciPy 1.0 Contributors}}]{Scipy_20_NMeth}
Virtanen, P., Gommers, R., Oliphant, T.~E., {et~al.} 2020, Nature Methods, 17, 261

\bibitem[{{Walch} {et~al.}(2013){Walch}, {Whitworth}, {Bisbas}, {W{\"u}nsch}, \& {Hubber}}]{Walch_13_MNRAS}
{Walch}, S., {Whitworth}, A.~P., {Bisbas}, T.~G., {W{\"u}nsch}, R., \& {Hubber}, D.~A. 2013, \mnras, 435, 917

\bibitem[{{Whitworth} {et~al.}(1994){Whitworth}, {Bhattal}, {Chapman}, {Disney}, \& {Turner}}]{Whitworth_94_AA}
{Whitworth}, A.~P., {Bhattal}, A.~S., {Chapman}, S.~J., {Disney}, M.~J., \& {Turner}, J.~A. 1994, \aap, 290, 421

\bibitem[{{Wright} \& {Mamajek}(2018)}]{Wright_Mamajek_18_MNRAS}
{Wright}, N.~J. \& {Mamajek}, E.~E. 2018, \mnras, 476, 381

\bibitem[{{Zamora-Avil{\'e}s} \& {V{\'a}zquez-Semadeni}(2014)}]{Zamora_Vazquez_14_ApJ}
{Zamora-Avil{\'e}s}, M. \& {V{\'a}zquez-Semadeni}, E. 2014, \apj, 793, 84

\bibitem[{{Zavagno} {et~al.}(2006){Zavagno}, {Deharveng}, {Comer{\'o}n}, {Brand}, {Massi}, {Caplan}, \& {Russeil}}]{Zavagno_06_AA}
{Zavagno}, A., {Deharveng}, L., {Comer{\'o}n}, F., {et~al.} 2006, \aap, 446, 171

\bibitem[{{Zucker} {et~al.}(2022){Zucker}, {Goodman}, {Alves}, {Bialy}, {Foley}, {Speagle}, {Gro{\^I}{\texttwosuperior}schedl}, {Finkbeiner}, {Burkert}, {Khimey}, \& {Swiggum}}]{Zucker_22_Natur}
{Zucker}, C., {Goodman}, A.~A., {Alves}, J., {et~al.} 2022, \nat, 601, 334

\end{thebibliography}


\begin{appendix}
\section{Properties of chain clusters}\label{app:prop_tables}

    In Table~\ref{tab:CrA}, we list the properties of all clusters part of the cluster chains discussed in this paper. This includes the number of stars considering different selection criteria, heliocentric Galactic Cartesian coordinates, distance from Earth, heliocentric Galactic Cartesian velocities, and the age of each cluster. Cluster speed measured relative to that of the progenitor clusters and that of the stars within clusters in Sco-Cen older than \SI{15}{\mega\year} (referred to as SC-15) are listed in Table~\ref{tab:relSpeed}. We characterize the properties of the group and their respective deviations using the median and median absolute deviation (MAD), respectively, due to their robustness to outliers. 
    
    We are using two reference systems for the cluster motion because we cannot uniquely attribute the origin of the feedback forces to the progenitor clusters of each chain with certainty. Therefore, we compute the cluster speed relative to the motion of the oldest progenitor clusters of each chain ($\phi$~Lup and $\sigma$~Cen), and the average motion of stars in Sco-Cen's clusters older than \SI{15}{\mega\year} (SC-15), as possible contributing feedback sources. SC-15 also includes the progenitor clusters of each cluster chain. In Fig.~\ref{fig:speed_time}, we show the cluster speed relative to the motion of the progenitor clusters as reference frame. To estimate the uncertainty of the relative cluster speed and that of Sco-Cen's average motion, we applied a bootstrap procedure where we sampled 1~000 realizations of cluster members with replacement. This generated distributions for each velocity component $(UVW)_{\text{i}}$ of the clusters along each chain, their progenitor clusters, and SC-15. We then subtracted the velocity vector distribution of the progenitor clusters or SC-15 from the distribution of velocity vectors of the clusters along each chain. This was done by (one-to-one) matching the 1~000 realizations from the reference velocity to that of each cluster in the chains. Subsequently, we computed the relative speed as the vector modulus of the differences ($|(UVW)_{\text{i}} - (UVW)_{\text{REF, i}}|$) as the relative speed for each relative velocity vector. Finally, the average relative cluster speed and the respective uncertainty, as given in Fig.~\ref{fig:speed_time} and Table~\ref{tab:relSpeed}, are the median and MAD of the resulting distribution. 

    
    The average motion of SC-15 and the respective uncertainties are the median and MAD of the distribution for each velocity component of a sampled velocity vector $UVW_{\text{SC-15, i}}$. This results in $UVW_{\text{SC-15}}$~=~(-6.4 $\pm$ 1.7, -19.7 $\pm$ 1.1, -5.5 $\pm$ 0.8) \si{\km\per\s}.
    
    \begin{table}[]
    \centering
    \caption{The relative cluster speed for all clusters in the CrA, LCC, and Upper-Sco chains.}
    \label{tab:relSpeed}
    \resizebox{0.84\columnwidth}{!}{%
        \begin{tabular}{cc|cc}
        \toprule \toprule
        \multicolumn{2}{l|}{\multirow{2}{*}{cluster chains}}& \multicolumn{2}{c}{$|\text{UVW}|_{\text{rel}}$  [km/s]} \\
                                        &                   & SC-15                  & PrC                       \\
        \midrule
        \multirow{6}{*}{CrA}            & $\phi$ Lup        & 0.55 $\pm$ 0.18        & —                         \\
                                        & $\eta$ Lup        & 0.93 $\pm$ 0.33        & 0.91 $\pm$ 0.32           \\
                                        & V1062-Sco         & 2.38 $\pm$ 0.25        & 2.13 $\pm$ 0.31           \\
                                        & Sco-Sting         & 1.84 $\pm$ 0.47        & 1.92 $\pm$ 0.52           \\
                                        & CrA-North         & 3.27 $\pm$ 0.30        & 3.42 $\pm$ 0.34           \\
                                        & CrA-Main          & 4.84 $\pm$ 0.18        & 4.99 $\pm$ 0.25           \\
        \midrule
        \multirow{5}{*}{LCC}            & $\sigma$ Cen      & 3.28 $\pm$ 0.20        & —                         \\
                                        & Acrux             & 4.36 $\pm$ 0.19        & 1.95 $\pm$ 0.18           \\
                                        & Musca-fg          & 5.79 $\pm$ 0.31        & 3.70 $\pm$ 0.43           \\
                                        & $\eta$ Cham       & 8.17 $\pm$ 0.36        & 5.87 $\pm$ 0.33           \\
                                        & $\epsilon$ Cham   & 6.98 $\pm$ 0.65        & 4.86 $\pm$ 0.72           \\
        \midrule
        \multirow{10}{*}{Upper-Sco}     & $\phi$ Lup        & 0.55 $\pm$ 0.18        & —                         \\
                                        & $\eta$ Lup        & 0.93 $\pm$ 0.33        & 0.91 $\pm$ 0.32           \\
                                        & Sco-Body          & 3.86 $\pm$ 0.24        & 3.88 $\pm$ 0.26           \\
                                        & $\rho$ Sco        & 2.48 $\pm$ 0.15        & 2.29 $\pm$ 0.17           \\
                                        & Antares           & 3.96 $\pm$ 0.13        & 3.90 $\pm$ 0.15           \\
                                        \cmidrule{3-4}
                                        & $\sigma$ Sco      & 2.89 $\pm$ 0.13        & 2.88 $\pm$ 0.17           \\
                                        & $\delta$ Sco      & 4.10 $\pm$ 0.11        & 4.21 $\pm$ 0.19           \\
                                        & $\beta$ Sco       & 4.44 $\pm$ 0.13        & 4.41 $\pm$ 0.16           \\
                                        & $\nu$ Sco         & 5.52 $\pm$ 0.11        & 5.62 $\pm$ 0.19           \\
                                        \cmidrule{3-4}
                                        & $\rho$ Oph        & 5.83 $\pm$ 0.11        & 5.96 $\pm$ 0.19           \\
        \bottomrule
        \end{tabular}
    }
    \tablefoot{The median and MAD of the cluster speed relative to the oldest progenitor clusters (PrC) of each cluster chain, $\phi$~Lup and $\sigma$~Cen, and relative to the motion of stars in Sco-Cen's clusters older than \SI{15}{\km\per\s} (SC-15). For details on the computation, see the text in Appendix~\ref{app:prop_tables}.}
    \end{table}

    \NewDocumentCommand{\anote}{}{\makebox[0pt][l]{$^{\ast}$}}
    \begin{table*}[]
    \centering
    \caption{Summary of the positional and kinematic properties for all clusters within the CrA, LCC, and Upper-Sco cluster chains.}
    \label{tab:CrA}
    \resizebox{\textwidth}{!}{%
    \begin{tabular}{cl|ccc|ccc|c|ccc|c}
    \toprule \toprule
    \multicolumn{2}{l|}{\multirow{2}{*}{cluster chains}}        & \multicolumn{3}{c|}{number of stars} & X                 & Y                & Z                     & dist. & U                & V                 & W                     & age   \\
                                             &                  & total     & RVs       & final        & \multicolumn{3}{c|}{heliocentric Gal. Cartesian coord. [pc]} & [pc]  & \multicolumn{3}{c|}{heliocentric Gal. Cartesian vel. [km/s]} & [Myr] \\
    \midrule
    \multicolumn{1}{c}{
    \multirow{6}{*}{CrA}}                    & $\phi$ Lup       & 1114      & 455       & 53\anote     & 112.4 $\pm$ 9.2   & -54.2 $\pm$ 6.8  & 40.3 $\pm$ 5.9        & 131   & -5.5 $\pm$ 1.3   & -19.6 $\pm$ 0.7   & -5.5 $\pm$ 1.0        & 16.9  \\
                                             & $\eta$ Lup       & 769       & 360       & 45\anote     & 124.6 $\pm$ 7.0   & -46.7 $\pm$ 8.8  & 24.7 $\pm$ 6.0        & 136   & -4.7 $\pm$ 0.9   & -20.0 $\pm$ 1.0   & -5.2 $\pm$ 0.5        & 15.3  \\
                                             & V1062-Sco        & 1029      & 309       & 21\anote     & 167.9 $\pm$ 4.5   & -51.3 $\pm$ 5.8  & 14.4 $\pm$ 2.7        & 177   & -3.6 $\pm$ 1.0   & -19.6 $\pm$ 0.4   & -4.2 $\pm$ 0.2        & 15.0  \\
                                             & Sco-Sting        & 132       & 42        & 7            & 131.3 $\pm$ 7.2   & -22.3 $\pm$ 5.3  & -7.5 $\pm$ 8.1        & 134   & -7.0 $\pm$ 1.5   & -18.6 $\pm$ 1.0   & -5.3 $\pm$ 0.3        & 14.5  \\
                                             & CrA-North        & 351       & 134       & 11\anote     & 144.7 $\pm$ 3.8   & -2.5 $\pm$ 2.7   & -36.2 $\pm$ 4.3       & 149   & -5.6 $\pm$ 0.8   & -17.3 $\pm$ 0.5   & -7.8 $\pm$ 0.7        & 11.6  \\
                                             & CrA-Main         & 96        & 56        & 29\anote     & 147.5 $\pm$ 2.2   & -0.3 $\pm$ 0.7   & -46.9 $\pm$ 1.1       & 155   & -4.3 $\pm$ 0.6   & -17.5 $\pm$ 0.6   & -9.6 $\pm$ 0.6        & 8.5   \\
    \midrule
    \multicolumn{1}{c}{
    \multirow{5}{*}{LCC}}                    & $\sigma$ Cen     & 1805      & 822       & 99\anote     & 59.9 $\pm$ 8.0   & -96.1 $\pm$ 7.1   & 17.1 $\pm$ 7.6        & 115   & -8.5 $\pm$ 1.0   & -20.7 $\pm$ 1.1  & -6.2 $\pm$ 0.8         & 15.5  \\
                                             & Acrux            & 394       & 190       & 24\anote     & 53.6 $\pm$ 2.3   & -91.5 $\pm$ 3.5   & -3.6 $\pm$ 3.4        & 106   & -9.3 $\pm$ 0.6   & -20.0 $\pm$ 0.8  & -7.7 $\pm$ 0.2         & 11.2  \\
                                             & Musca-fg         & 95        & 44        & 8            & 51.8 $\pm$ 2.1   & -86.5 $\pm$ 2.6   & -18.2 $\pm$ 2.6       & 102   & -10.2 $\pm$ 0.7  & -18.7 $\pm$ 1.8  & -8.7 $\pm$ 0.4         & 10.2  \\
                                             & $\eta$ Cham      & 30        & 14        & 4            & 34.9 $\pm$ 0.4   & -84.7 $\pm$ 1.0   & -36.2 $\pm$ 1.8       & 99    & -11.7 $\pm$ 0.7  & -19.3 $\pm$ 0.2  & -10.6 $\pm$ 0.4        & 9.4   \\
                                             & $\epsilon$ Cham  & 39        & 20        & 4            & 49.8 $\pm$ 1.0   & -84.6 $\pm$ 1.7   & -27.9 $\pm$ 1.0       & 102   & -10.2 $\pm$ 0.7  & -19.7 $\pm$ 0.7  & -10.5 $\pm$ 1.1        & 8.8   \\
    \midrule
    \multicolumn{1}{c}{
    \multirow{10}{*}{Upper-Sco}}             & $\phi$ Lup       & 1114      & 455       & 53\anote     & 112.4 $\pm$ 9.2   & -54.2 $\pm$ 6.8  & 40.3 $\pm$ 5.9          & 131   & -5.5 $\pm$ 1.3   & -19.6 $\pm$ 0.7   & -5.3 $\pm$ 1.0      & 16.9  \\
                                             & $\eta$ Lup       & 769       & 360       & 45\anote     & 124.6 $\pm$ 9.8   & -46.7 $\pm$ 8.8  & 24.7 $\pm$ 6.0          & 136   & -4.7 $\pm$ 0.9   & -20.0 $\pm$ 1.0   & -5.2 $\pm$ 0.5      & 15.3  \\
                                             & Sco-Body         & 373       & 143       & 17\anote     & 137.7 $\pm$ 7.0   & -26.1 $\pm$ 5.6  & 17.3 $\pm$ 10.5         & 141   & -3.9 $\pm$ 1.4   & -16.8 $\pm$ 0.6   & -7.6 $\pm$ 0.5      & 14.7  \\
                                             & $\rho$ Sco       & 240       & 108       & 48\anote     & 129.2 $\pm$ 5.2   & -24.5 $\pm$ 7.2  & 42.9 $\pm$ 3.7          & 139   & -4.3 $\pm$ 1.1   & -18.1 $\pm$ 0.4   & -4.0 $\pm$ 0.5      & 13.7  \\
                                             & Antares          & 502       & 270       & 131\anote    & 131.7 $\pm$ 5.9   & -16.6 $\pm$ 4.0  & 41.3 $\pm$ 3.8          & 139   & -3.9 $\pm$ 1.1   & -16.0 $\pm$ 0.7   & -5.8 $\pm$ 0.5      & 12.7  \\
                                             \cmidrule{3-13}
                                             & $\sigma$ Sco     & 544       & 263       & 158\anote    & 149.3 $\pm$ 4.9   & -23.6 $\pm$ 3.1  & 49.0 $\pm$ 3.2          & 159   & -4.2 $\pm$ 1.0   & -17.2 $\pm$ 0.7   & -6.2 $\pm$ 0.5      & 10.0  \\
                                             & $\delta$ Sco     & 691       & 451       & 268\anote    & 129.7 $\pm$ 2.8   & -22.5 $\pm$ 4.3  & 53.7 $\pm$ 3.0          & 142   & -6.2 $\pm$ 0.7   & -16.2 $\pm$ 0.5   & -7.6 $\pm$ 0.6      & 9.8   \\
                                             & $\beta$ Sco      & 285       & 186       & 107\anote    & 140.8 $\pm$ 3.8   & -16.6 $\pm$ 2.5  & 61.0 $\pm$ 3.2          & 154   & -3.0 $\pm$ 0.8   & -16.2 $\pm$ 0.4   & -6.8 $\pm$ 0.6      & 7.6   \\
                                             & $\nu$ Sco        & 150       & 102       & 70\anote     & 127.9 $\pm$ 2.4   & -12.3 $\pm$ 1.3  & 54.2 $\pm$ 1.3          & 139   & -5.2 $\pm$ 0.6   & -15.4 $\pm$ 0.4   & -9.0 $\pm$ 0.4      & 5.8   \\
                                             \cmidrule{3-13}
                                             & $\rho$ Oph       & 535       & 281       & 155\anote    & 131.6 $\pm$ 2.8   & -15.8 $\pm$ 1.0  & 40.9 $\pm$ 1.9          & 139   & -5.7 $\pm$ 0.9   & -15.2 $\pm$ 0.6   & -9.2 $\pm$ 0.8      & 3.8   \\
    \bottomrule
    \end{tabular}%
    }
    \tablefoot{Columns 3-5 list the number of stars per cluster, the number of radial velocities, and the final number of radial velocities after applying the selection criteria, including the radial velocity uncertainty cut at \SI{1}{\km\per\s} and sigma-clipping of the radial velocity distribution. Clusters that underwent sigma-clipping are marked with an asterisk ($^{\ast}$). See Sect.~\ref{sec:data} for details. Columns 6-8 list the median and median absolute deviation (MAD) of each cluster's position in heliocentric Galactic Cartesian coordinates. The distance to each cluster is given in column 8 and is sourced directly from \citetalias{Ratzenboeck_23a_AA}. Columns 10-12 list the median and MAD of each velocity component in heliocentric Galactic Cartesian coordinates. The cluster ages estimated by \citetalias{Ratzenboeck_23b_AA} are given in column 13.}
    \end{table*}

    \begin{table*}[]
    \centering
    \caption{List of parameters resulting from the linear regression of the relative speed-age relation in Fig.~\ref{fig:speed_time}, with respect to the progenitor clusters (PrC) within each cluster chain, as well as SC-15, and the mass-age relation shown in Fig.~\ref{fig:mass_time}.}
    \label{tab:param}
    \resizebox{0.85\textwidth}{!}{%
        \begin{tabular}{l|cccc|cc}
        \toprule \toprule
                                    & \multicolumn{2}{c}{average acceleration rel. to PrC} & \multicolumn{2}{c|}{average acceleration rel. to SC-15} & \multicolumn{2}{c}{Mass depletion} \\
                                    & slope             & intercept                        & slope            & intercept                            & slope          & intercept         \\
                                    & [km/s/Myr]        &  [km/s]                          & [km/s/Myr]       &  [km/s]                              & [\#/Myr]        & [\#]             \\
        \midrule
        CrA                         & 0.57 $\pm$ 0.06   & 10.48 $\pm$ 0.84                 & 0.49 $\pm$ 0.08  & 9.11 $\pm$ 1.09                      & -137 $\pm$ 23  & -1293 $\pm$ 320   \\
        \midrule
        LCC                         & 0.75 $\pm$ 0.12   & 11.60 $\pm$ 1.33                 & 0.63 $\pm$ 0.15  & 12.07 $\pm$ 1.72                     & -266 $\pm$ 21  & -2457 $\pm$ 244   \\
        $\quad$ excl. $\eta$~Cham   & 0.69 $\pm$ 0.12   & 10.53 $\pm$ 1.40                 & 0.53 $\pm$ 0.11  & 10.53 $\pm$ 1.32                     &                &                   \\
        \midrule
        Upper-Sco                   & 0.39 $\pm$ 0.05   & 7.76 $\pm$ 0.70                  & 0.32 $\pm$ 0.05  & 7.51 $\pm$ 0.58                      & -39 $\pm$ 12   & 73 $\pm$ 143      \\
        $\quad$ Upper-Sco~I         & 0.69 $\pm$ 0.15   & 12.37 $\pm$ 2.23                 & 0.69 $\pm$ 0.16  & 13.01 $\pm$ 2.33                     & -175 $\pm$ 26  & -1987 $\pm$ 405   \\
        $\quad$ Upper-Sco~II        & 0.55 $\pm$ 0.16   & 8.84 $\pm$ 1.37                  & 0.47 $\pm$ 0.09  & 8.56 $\pm$ 0.74                      & -134 $\pm$ 20  & -703 $\pm$ 167    \\
        \bottomrule
        \end{tabular}
    }
    \end{table*}

    \subsection{Acceleration estimate for the cluster chains}\label{saap:acceleration}

        Assuming the sequential formation of clusters from the same progenitor molecular cloud, we expect to observe an increase of speed over time along the cluster chain.
        To estimate this average acceleration for each cluster chain, we applied a Bayesian linear regression to the data points in Fig.~\ref{fig:speed_time}, where $x$ denotes the time when the cluster was formed and $y$ denotes the relative cluster speed. We adopt the Bayesian framework because it yields a distribution for each of the fitting parameters instead of point estimates, which gives a meaningful uncertainty for the prediction of the acceleration. 
        This was done using the Python library \texttt{PyMC} \citep{PyMC_16} with a simple linear model $f(x) = ax + b$, where $a$ is the slope and $b$ is the intercept of the relation. Eventually, $a$ gives the acceleration in units of \si{\km\per\s\per\mega\year}, and $b$ is the extrapolated speed for a cluster born today ($x$ = \SI{0}{\mega\year}). We chose priors according to the average parameters of linear Least-Squares fits previously applied to the data. We chose a Normal distribution centered on \SI{12}{\km\per\s} with a standard deviation of \SI{6}{\km\per\s} as the prior for the intercept and a Uniform distribution between 0 and \SI{1}{\km\per\s\per\mega\year} as the prior for the slope of the relation. We adopted the median and MAD of the resulting parameter distributions as the final fit parameters. All parameters are listed in Table~\ref{tab:param}.
        We confirmed that the linear fit to the data is preferred over a quadratic fit by using a model comparison, that gives a statistical preference for either of the fitting functions applied to the data.

    \subsection{Measuring the depletion in star formation rate}\label{saap:sfr}

        A linear fit to the mass relation reveals the trend in the star formation rate over time along each cluster chain. For estimating the average mass depletion, we again applied linear Bayesian regression curves to the data points in Fig.~\ref{fig:mass_time}. The priors were also based on the parameters of Least-Squares fits to each chain, which resulted in an average slope of \SI{-150}{\per\mega\year} and an intercept of \num{-1500} stars. A relatively uninformative prior was selected for the intercept, modeled as a Normal distribution centered on \num{-1500} stars with a standard deviation of \num{1500}. The prior for the slope was set to a Uniform distribution between \num{-300} and \SI{-100}{\per\mega\year}. The decrease in mass of the Upper-Sco chain deviates significantly from the rest of the chains. To fit it, we used priors of \num{-100\pm500} stars for the intercept, and a uniform distribution between \num{-100} and \SI{0}{\per\mega\year} for the slope. The final fit parameters were derived using the median and MAD of the resulting parameter distributions. The complete set of parameters is provided in Table~\ref{tab:param}. By multiplying the number of stars by an average stellar mass of \SI{0.42}{\solarmass}, as derived from the \cite{Kroupa_01_MNRAS} mass function, we estimated the change of star formation rate overtime. We computed an average depletion in number of stars of about \SI{-180}{\per\mega\year}, corresponding to an average decrease in star formation rate of about \SI{-80}{\solarmass\per\mega\year} along the cluster chains.

\section{Agreement of radial velocities in clusters and residual clouds}\label{saap:rv}
    Following the methods of \cite{Grossschedl_21_AA} and \citetalias{Posch_23_AA}, we validated the assumed kinematic connection by comparing radial velocity measurements of the clouds to the youngest clusters of each chain of clusters. Using CO data from \cite{Dame_01_ApJ}, we compared the radial velocity of the Ophiuchus cloud to that of $\rho$~Oph. The radial velocity of the cloud is determined by calculating the mean and standard deviation of the pixels at the location where the cloud overlaps with the youngest cluster. The converted radial velocities of the clusters are computed as the means and standard deviations of 1~000 samples drawn randomly from a Normal distribution, which is based on the means and standard deviations of the heliocentric measurements of the cluster members. 
    \cite{Dame_01_ApJ} provided radial velocity measurements relative to the local standard of rest (LSR), likely using (UVW)$_\mathrm{Sun}$~=~(10.0, 15.4, 7.8)~\si{\km\per\s} as solar motion \citep{Kerr_LyndenBell_86_HiA}. Consequently, we corrected our clusters' heliocentric radial velocity measurement using this definition of the LSR. In \citetalias{Posch_23_AA}, we found that the radial velocity of the head of the CrA molecular cloud (RV$_{\mathrm{LSR}}$~=~\SI{5.7\pm1.4}{\km\per\s}) matches that of the CrA~Main cluster (RV$_{\mathrm{LSR}}$~=~\SI{6.0\pm1.3}{\km\per\s}) within the uncertainties.    
    For $\rho$~Oph, we obtained RV$_{\mathrm{LSR}}$~=~\SI{3.6\pm1.9}{\km\per\s}, which aligns with the radial velocity measurement for the Ophiuchus molecular cloud from \cite{Dame_01_ApJ}, RV$_{\mathrm{LSR}}$~=~\SI{4.1\pm1.2}{\km\per\s}, within the uncertainties. For $\nu$~Sco, we computed RV$_{\mathrm{LSR}}$~=~\SI{3.7\pm1.6}{\km\per\s}, aligning with the radial velocity of CO gas in the B40 molecular cloud of RV$_{\mathrm{LSR}}$~=~\SI{3.3\pm0.5}{\km\per\s}.
    
    In the LCC chain, we compared the radial velocity of $\epsilon$~Cham with that of the Blue Cloud.
    There is a radial velocity signal of the cloud's dense molecular head in the \cite{Dame_01_ApJ} CO survey, measuring RV$_{\mathrm{LSR}}$~=~\SI{4.4\pm0.9}{\km\per\s}. This matches the measurement of \cite{Boulanger_98_AA} (\SI{4.8\pm0.9}{\km\per\s}) and the LSR-corrected radial velocity of $\epsilon$~Cham (\SI{4.3\pm1.0}{\km\per\s}) within the uncertainties. \cite{Nehme_08_AA} reported a slightly lower radial velocity for the Blue Cloud at \SI{2.8\pm0.2}{\km\per\s}, which falls outside this range. Neither \cite{Boulanger_98_AA} nor \cite{Nehme_08_AA} specified the LSR definition they used, so we assumed that they used the solar motion measurement of \citet{Kerr_LyndenBell_86_HiA}, probably common until 2010, when \citet{Schoenrich_10_MNRAS} published the current widely adopted solar motion measurement.

\section{Momentum analysis of the Ophiuchus molecular cloud}\label{app:momentum}

    We can approximate the momentum of a molecular cloud with its mass and velocity. The cloud velocity can be approximated as that of its associated young stars, based on the observation that young stars share the motion of their parent molecular cloud over several million years \citep{Tobin_09_ApJ, Hacar_16_AA, Grossschedl_21_AA}.
    The analysis outlined in this section was previously conducted for the CrA chain of clusters, as described in \citetalias{Posch_23_AA}. The results of the analysis of the CrA chain and those for the Upper-Sco chain are listed in Table~\ref{tab:chains}. We refrained from estimating the cloud momentum of the Blue Cloud at the tip of the LCC chain because the mass is poorly constrained for this low-density gas structure.

    \subsection{Cloud mass}\label{saap:mass}
    We estimated the cloud mass of the Ophiuchus molecular cloud using high-resolution 3D dust maps from \cite{Leike_20_AA} and \cite{Edenhofer_24a_AA}. We verified this approach by calculating the mass of the CrA molecular cloud and comparing it to existing estimates based on infrared extinction maps.
    We selected a box that included the gas structure and summed its HI density. Following the procedures in \cite{Bialy_21_ApJ} and \cite{ONeill_24_ApJ}, each voxel was converted to the total hydrogen nuclei density (N$_{\mathrm{HI}}$) by multiplying with \SI{880}{\per\cm} and \SI{1653}{\per\cm} for the 3D dust maps of \cite{Leike_20_AA} and \cite{Edenhofer_24a_AA}, respectively.
    Similarly to \cite{ONeill_24_arXiv} and \cite{Edenhofer_24b_AA}, we computed the gas mass in units of \si{\kg}:

    \begin{equation*}
        M_{\mathrm{HI}} = \mu \, m_{\mathrm{p}} \, C \; \sum_i N_{\mathrm{HI}, i} \, \mathrm{d}V_i
    \end{equation*}

    Here, $\mu$ denotes the mass correction factor considering the average abundance of atomic and molecular hydrogen and (\num{1.37}), $m_{\mathrm{p}}$ denotes the mass of a proton, C is the conversion factor of \si{\cubic\cm} to \si{\cubic\parsec} ((\num{3.086E18})$^3$), and $\mathrm{d}V_i$ is the unit volume of a voxel within the box in units of \si{\cubic\parsec}.
    
    To verify this method, we applied it to the CrA molecular cloud using only the \cite{Leike_20_AA} map and compared the result with the mass derived by \cite{Alves_14_AA, Posch_23_AA}. Using a box volume defined by X = [125, 165]~\si{\parsec}, Y = [-15, 20]~\si{\parsec}, and Z = [-75, -35]~\si{\parsec}, we calculated a mass of approximately \SI{4000}{\solarmass}, which is lower than the measurements reported in the literature, \SIrange{6000}{9000}{\solarmass}.
    Applying the method to the Ophiuchus molecular cloud, the dimensions of the box are X = [120, 150], Y = [-20, 10], and Z = [20, 50] and we obtained a mass of approximately \SI{11000}{\solarmass} for both 3D dust maps.
    Based on the method verification with the CrA cloud, we considered the cloud mass estimate for the Ophiuchus cloud to be a conservative lower limit. We adopted a Gamma distribution for the cloud mass, having an extended tail towards higher masses. The mass distribution of 10~000 estimates peaks at the derived cloud mass of \SI{11000}{\solarmass}, with the 2.5 and 97.5 percentiles at \SI{10000}{\solarmass} and \SI{18000}{\solarmass}, respectively.
    
    \subsection{Estimating cloud momentum}\label{saap:momentum}
    Multiplying the cloud mass by the cloud's velocity provides an estimate of the cloud's momentum ($P = M_{\mathrm{cloud}} \cdot v_{\mathrm{cloud}}$). We adopted the velocity of $\rho$~Oph relative to both reference frames as the cloud's velocity ($\sim$~\SI{6}{\km\per\s}), since the radial velocity of $\rho$~Oph aligns with that of the Ophiuchus cloud. Bootstrapping 10~000 times over the cluster members with replacement yields a Normal distribution for the cloud velocity. By multiplying the mass and the velocity distributions, we derived a median cloud momentum of $\sim$~\SI{70000}{\km\per\s\solarmass}, with the 2.5 and 97.5 percentiles at \SIrange{60E3}{100E3}{\km\per\s\solarmass}.

    Assuming that this cloud momentum was primarily influenced by stellar feedback from the OB association, we can approximate the required amount of feedback. For simplicity, we considered only the feedback from SN explosions, as they are likely the dominant force. \citetalias{Posch_23_AA} estimated that stellar winds have a minor impact, and photoionization possibly also contributes only slightly (see Sect.~\ref{sub:formation}). We assumed a Uniform distribution for the radius of the progenitor cloud between \SIrange{8}{10}{\parsec}, based on the current radius of the 3D dust distribution of the Ophiuchus molecular cloud as measured in the Leike map \citep{Leike_20_AA} ($\sim$~\SIrange{4}{5}{\parsec}). 
    Our traceback analysis shows that most clusters were closer than \SI{20}{\parsec} and some closer than \SI{15}{\parsec} during the onset of star formation in $\rho$~Oph (see Fig.~\ref{fig:rhoOph} for the three closest clusters). Therefore, we adopted a Normally distributed distance between the youngest clusters and the feedback source of \SI{15\pm5}{\parsec}. We again used 10~000 realizations for these distributions. Following the methods outlined in \citetalias{Posch_23_AA}, we then computed the fraction of SN momentum impacting the cloud surface and derived a distribution for the number of SN required to account for the cloud momentum. A median number of three SN explosions is sufficient to account for the current momentum of the Ophiuchus molecular cloud, with the 2.5 and 97.5 percentiles at \num{0.4} and \num{9.1} SNe.

\section{Orbital traceback of $\rho$~Oph}\label{app:rhoOph}
    For integrating cluster orbits relative to the LSR, we used the \texttt{galpy} Python library \citep{galpy_15} with the default parameters. The default parameters include \texttt{MWPotential2014} as the Galactic potential, solar motion relative to the LSR of \cite{Schoenrich_10_MNRAS} ($UVW_{\mathrm{\odot}, LSR}$) = (-11.1, 12.24, 7.25) \si{\km\per\s}, and the position of the Sun relative to the Galactic center ($XYZ_{\mathrm{G}}$)~=~(8122.0, 0.0, 20.8) \si{\parsec} \citep{GravityColl_18_AA, Bennett_Bovy_19_MNRAS}.
    We estimated the average positions and velocities through bootstrapping with replacement, using 1~000 realizations. From the median phase space information of each realization, we integrated the orbits \SI{20}{\mega\year} back in time, and obtained a distribution of orbits for all clusters.
    
\begin{figure}
        \centering
        \includegraphics[width=\columnwidth]{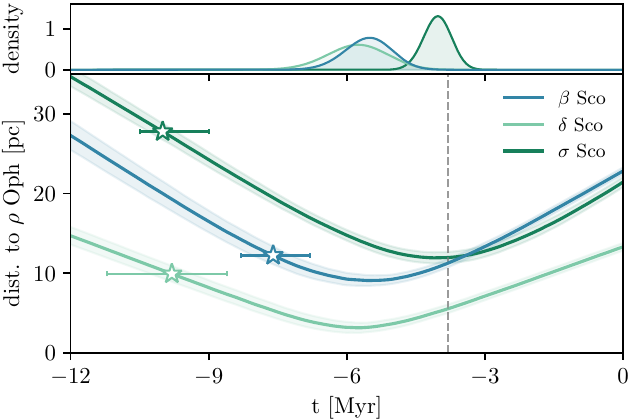}
        \caption{Euclidean distance between $\rho$~Oph and each of the clusters $\sigma$~Sco, $\delta$~Sco, and $\beta$~Sco over the past \SI{12}{\mega\year}. Ages and their corresponding uncertainties for $\sigma$~Sco, $\delta$~Sco, and $\beta$~Sco are indicated with a star symbol (taken from \citetalias{Ratzenboeck_23b_AA} and given in Table~\ref{tab:CrA}). The vertical gray dashed line indicates the age of $\rho$~Oph. Prior to star formation, the motion of these clusters corresponds to that of their primordial clouds. In the upper panel, we show the distribution of times of minimum distance.}
        \label{fig:rhoOph}
    \end{figure}

    Then, we calculated the Euclidean distances between $\rho$~Oph and the clusters $\sigma$~Sco, $\delta$~Sco, and $\beta$~Sco over the past \SI{12}{\mega\year} for each integrated orbit. Figure~\ref{fig:rhoOph} shows the average distances with the according deviation. In the bottom panel of the figure, we show the median orbital distances to $\rho$~Oph and its respective uncertainty estimation using the MAD. In the upper panel, we show the distribution of times of minimum distance. The average minimum distances and the corresponding average times of the minimum distance are listed in Table~\ref{tab:rhoOph} for each of the three clusters.

    As part of this investigation, we found that approximately \SI{6}{\mega\year} ago, $\rho$~Oph was located within \SI{3\pm3}{\parsec} of the $\delta$~Sco cluster, within \SI{9\pm2}{\parsec} of the $\beta$~Sco cluster, and about \SI{4}{\mega\year} ago, within \SI{12\pm1}{\parsec} of the $\sigma$~Sco cluster. This confirms previous results from \cite{MiretRoig_22_AA}. Given that these three clusters host massive stars, the formation of $\rho$~Oph shortly after the encounter supports a scenario in which stellar feedback induced the star formation process in this cluster.

    \begin{table}[]
    \centering
    \caption{Summary table of minimum distances and times at which the minimum distances occurred between $\rho$~Oph and the clusters $\beta$~Sco, $\delta$~Sco, and $\sigma$~Sco.}
    \label{tab:rhoOph}
    \resizebox{0.67\columnwidth}{!}{%
        \begin{tabular}{ccc}
        \toprule \toprule
        distance to  & d$_{\mathrm{min}}$ [pc] & t(d$_{\mathrm{min}}$)  [Myr] \\
        \midrule
        $\beta$~Sco  & 9.1 $\pm$ 0.7           & -5.5 $\pm$ 0.2               \\
        $\delta$~Sco & 3.1 $\pm$ 0.7           & -5.7 $\pm$ 0.3               \\
        $\sigma$~Sco & 11.9 $\pm$ 0.7          & -4.1 $\pm$ 0.1               \\
        \bottomrule
        \end{tabular}
    }
    \tablefoot{From the orbits calculated using 1~000 bootstrapped samples, we derive distributions of minimum distances between the cluster orbits and the corresponding times of minimum distances. The reported values and uncertainties are the median and median absolute deviation of these distributions, respectively.}
    \end{table}
    
\end{appendix}

\end{document}